\documentclass[twocolumn]{emulateapj}

\usepackage{amsmath,natbib, appendix}
\usepackage{longtable}
\usepackage{graphicx}

\slugcomment{Submitted to ApJ}

\shorttitle{FRAGMENTATION IN THE FIRST GALAXIES}

\shortauthors{SAFRANEK-SHRADER ET AL.}

\begin{document}
\title{Fragmentation in the First Galaxies}

\author{Chalence Safranek-Shrader, 
Volker Bromm,
and Milo\v s Milosavljevi\'c}

\affil{Department of Astronomy and Texas Cosmology Center, University of Texas at Austin, Austin, TX 78712}

\newcommand{\htwo}{$\textrm{H}_2$}
\newcommand{\htwoeq}{\textrm{H}_2}

\begin{abstract}
Motivated by recent simulations of galaxy formation in which
protogalaxies acquire their baryonic content through cold
accretion, we study the gravitational fragmentation of cold streams
flowing into a typical first galaxy. We use a one-zone hydrodynamical model to 
examine the thermal evolution of the gas flowing into a $10^{8}\,M_{\odot}$ DM halo at redshift $z=10$. The goal is to gain an understanding of the expected fragmentation mass scale and thus the
characteristic mass of the first population of stars to form by shock
fragmentation. Our model accurately describes the chemical and thermal
evolution of the gas  as we are specifically concerned with how the chemical abundances and initial conditions of the low density, metal enriched, cold accretion streams that pass an accretion shock alter
the cooling properties and tendency to fragment in the post-shock gas.
Cold accretion flows are not shock heated at the virial radius but instead
flow along high-baryonic-density filaments of the cosmic web and penetrate deep into the
host halo of the protogalaxy. In this physical regime, if molecular cooling is absent because of a strong LW background, we find there to be a sharp drop in the fragmentation mass at a metallicity of $Z \sim10^{-4}\,Z_{\odot}$. If, however, $\text{H}_2$ and HD molecules are present, they dominate the cooling at $T < 10^{4}\, \text{K}$, and metallicitiy then has no effect on the fragmentation properties of the cold stream.  For a solar abundance pattern of metallicity, O is the most effective metal coolant throughout the evolution, while for a pair instability supernova (PISN) metallicity yield, $\text{Si}^{+}$ is the most effective coolant. PISN abundance patterns also exhibit a slightly smaller critical metallicity. Dust grains are not included in our chemical model, but we argue their inclusion would not significantly alter the results. We also find that this physical scenario allows for the formation of stellar clusters and large, $10^{4}\, M_{\odot}$ bound fragments, possibly the precursors to globular clusters and supermassive black holes. Lastly, we conclude that the usual assumption of isobaricity for galactic shocks breaks down in gas of a sufficiently high metallicity, suggesting that metal cooling leads to thermal instabilities.

\end{abstract}


\section{Introduction}
An important open question in cosmology is to identify the conditions
under which the first Population II (Pop II) stars could have formed.
It has been argued that clusters of predominantly low-mass stars
could have arisen in the first galaxies at $z=10$ \citep{BC02, GB06}. Their
emergence straddles the end of the so-called cosmic dark ages, arguably the most uncharted time in the history of the
universe \citep{BarLoeb01, MirES03}. Between the release of the cosmic microwave background (CMB) at $z\simeq 1,100$ and the highest redshift gamma ray burst (GRB) at $z\simeq8.2$ \citep[][]{Salvaterra:09,Tanvir:09}, no direct observations have been made yet, thus rendering theoretical models our best tools for exploring this formative cosmic epoch \citep[for indirect probes of this era, see][]{GIS00, CHR02, CEN03, FRE05, FRE09}.

 The standard cosmological model ($\Lambda$CDM) predicts the first stars (Pop III) formed at redshifts $z\simeq20-30$ in $\sim10^{6}\,M_{\odot}$ dark matter `minihalos' \citep[e.g.,][]{CR86, HAI96, TEG97} and this brought about an end to the cosmic dark ages.
The cooling properties of
$\text{H}_{2}$, the most effective primordial gas coolant at
temperatures below $10^{4} \text{ K}$, set a characteristic
mass scale of $100~M_\odot$ for Pop III stars \citep{ABN00, BCL02}. The
primordial initial mass function (IMF) would then have been top-heavy. This result
has been confirmed by subsequent studies \citep[e.g.,][]{OP2003, ONor0700, Y06, Y08}. Further investigations were directed at elucidating their properties and key implications, such as
the chemical signature \citep{HW02, Iwamoto05}, the detailed IMF \citep{NU01}, the effect on reionization \citep{WL03, SOK04}, the
formation rate \citep{TS09}, and the tendency toward binarity \cite[][]{TUR09, STA09}. These massive Pop~III stars had extremely hot photospheres \citep[$T_{\rm eff}
\sim 10^5\,\text{K}$,  see][]{BK01, SCH02}, and very short lifetimes
of $\sim$ 3 Myr \citep[for the case against this standard picture see][]{TUM04, SL06, CLA10}.

It is still not known with any certainty when and how the
universe shifted its star formation mode from the top-heavy IMF for Pop III to the
roughly Salpeter distribution seen today. \citet{BRO01} suggested it is the chemical input from Pop III
stars that allowed gas in minihalos halos to cool more efficiently and fragment at a
smaller mass scale. They found that when molecular cooling is excluded by a soft-UV photodissociating background,
gas enriched to $Z_{\rm crit} \simeq 10^{-3.5}\,Z_{\odot}$, the so-called critical metallicity, fragments into smaller mass clumps. The idea
of metallicity being the main driver of the mode transition was advanced by many other studies. \citet{SS06} found individual critical metallicities for C, Fe, Si, and O and suggested that for densities and temperatures conducive to Pop III star formation, these species contribute more to the cooling rate than atomic hydrogen or $\text{H}_2$. \citet{SMI09} found that there exists a critical metallicity window, due
in part to the CMB temperature floor, which regulates the transition. In a series of papers \citep{OMU00, OMU05,OMU08} dust was shown to have a significant impact on the Pop III/II transition. Its inclusion lowers the critical metallicity derived from pure metal line cooling by several orders of magnitude. Dust grains act as a catalyst for molecule formation and radiate thermally, however these effects generally become important late in the pre-stellar gas contraction phase when density is high. \citet{SO09} argued that the interplay of dust and metal cooling is the key driver, together with the CMB temperature floor.
A different view was developed in
a series of studies \citep{JAP07, Jappsen:09a,JAP09}, now taking into account molecular cooling. According to these authors, metallicity plays no role in the fragmentation of collapsing
clouds while initial conditions, set by the details of galaxy formation, the CMB temperature, turbulence, and rotation, dominate the Pop. III/II transition.  This star formation mode transition is clearly a complex, gradual process, regulated
in part by all of these factors.

Most studies of the critical metallicity
to date have focused on $\sim 10^{6} \,M_\odot$ DM halos, or minihalos, as the primordial star formation environment. However, the majority of the first Pop II stars
likely formed in deeper potential wells, possibly in halos
that can cool via atomic hydrogen cooling \citep[e.g.,][]{Betal09}. Whereas the quasi-hydrostatic, roughly spherical nature of baryonic collapse inside of minihalos admits
modeling of low metallicity cooling in a rather straightforward
fashion, the hydrodynamics inside the first galaxies is much
more complex. Simulations with realistic cosmological initial conditions have recently demonstrated how cold accretion streams feed dense, turbulent gas to the center of the galaxy, where second generation star formation will take place \citep[e.g.,][]{GB08,WA07,WA08}. Since detailed simulations of the fragmentation properties in such a cloud
with different levels of pre-enrichment are still lacking, it is
timely to carry out an exploratory survey of the crucial physics involved
in the fragmentation of such shocked, stream-fed gas.

These cosmic filaments play a pivotal role in galaxy formation by adding a 
third accretion mode to the two processes that have traditionally been thought to partake in galaxy formation. 
Spheroidal galaxy components were thought to have been built by gas channeling in galactic mergers \citep{TOO77, WR78} and disk components by accretion from a shock-heated
intergalactic medium (IGM) during virialization \citep{RO77, SIL97}. Hot accretion occurs when gas is
shock heated to approximately the virial temperature close to the virial radius of the DM halo; the gas then cools, falls inward, and circularizes in the galactic disk. Recently, however, a new mode of
accretion has been found to dominate in high-$\sigma$ cosmological peak halos at high
redshift \citep{DB06,DEK09}. This mode, coined `cold flow accretion,' occurs along
cosmic filaments that feed gas directly to the inner
parts of a galaxy, leading to shocks at radii much smaller than the halo virial radius \citep[see, also,][]{WA07,Binney:04}. This differs qualitatively from hot accretion in
that the infalling gas passes through the virial shock unaffected, has a much
higher density, and is deposited much closer to the center of the
galaxy. Cold accretion is theoretically very appealing. It presents a solution to the observed blue-red galaxy
bimodality \citep{BAL04}. It provides a natural explanation
for high redshift galaxies with large star formation rates (SFRs) but no evidence of a major
merger \citep[e.g.,][]{Genzel:06}.  Also, cold accretion
naturally arises in many simulations of galaxy formation \citep{KER05, DB06, WA07, GB08,
HAR08,Regan:09}, suggesting this type of accretion is natural to high-$\sigma$ peak halos at high redshifts within $\Lambda$CDM models.

DM halos with masses $\sim 10^{8} \,M_\odot$ represent the
smallest structures that can be classified as `first galaxies' \citep{KY05, REA06, WA07, GB08, Betal09}.
Halos of this mass have virial temperatures $T_{\text{vir}} >  10^{4} \,\text{K}$ which allows
efficient cooling via excitation of atomic hydrogen \citep[e.g.,][]{Haiman:97,TEG97,MiraldaEscude:98,Barkana:99,Oh:99}. This is the halo mass scale where we first expect to see sustained, possibly
self-regulated, star formation, similar to the present-day case.
Finally, in $10^{8}\,M_\odot$ halos, gas accreting along cosmic filaments will drive supersonic turbulence \citep{WA07, GB08,Regan:09,KLE10}, fundamentally changing the star formation dynamics \citep{PN02,MacLow:04, MOst07}. In this work, as in many others, we define a first galaxy to be one of the first DM halos to form with mass $\geq5\times10^{7}\,M_{\odot}$, thus meeting the requirement that gas heated to the virial temperature is able to cool through hydrogen excitation (or Lyman-$\alpha$) emission.

A further motivation for this work is understanding the formation of low-luminosity dwarf spheriodal satellite galaixes (dSph) and globular clusters (GCs). These structures have similar luminosities, stellar masses, generally no gas nor recent star formation, and contain metal-poor, old stellar populations. However, GCs are small ($\sim 1-10\,\text{pc} $) and do not seem to contain much DM while dSph galaxies are much larger ($>100\,\text{pc} $) and are heavily dominated by DM as indicated by their large mass-to-light ratios \citep{MAT98,STR08}. It is clear these are two structurally distinct populations \citep[e.g.,][]{KOR85,BEL07}, but it is necessary to understand the cosmological circumstances of their formation. To answer whether the first galaxies are progenitors to dSph galaxies, GCs, or both requires an understanding of star formation in these objects.

With this in mind, we investigate the fragmentation properties of the metal-pre-enriched cold flow accretion in protogalaxies at high redshift, and explore implications for star formation in these objects.
Dense, metal-enriched baryonic streams flow along the filaments of the cosmic web and penetrate deep into a protogalaxy.  The multiple (e.g., ternary) streams collide with each other, or collide with a turbulently-supported gas accumulation that has already collected in the center of the protogalaxy.  Since the stream inflow is supersonic, the colliding gas must pass a strong shock, but then cools rapidly under approximate isobaric conditions.  We identify the characteristic fragmentation mass scale in the shocked, isobarically cooling gas.  We find that if a strong photodissociating Lyman-Werner (LW) background has significantly suppressed molecule formation, the characteristic mass scale depends strongly on the metallicity content of the gas.

The outline of this work is as follows. In Section \ref{sec:cooling} we describe our
methodology for the cooling of shocked, metal-enriched gas.  Section \ref{sec:initial_conditions} motivates the initial conditions for our model, including metal abundances and the physics of cold accretion. In Section \ref{sec:thermal_evolution} we discuss our fragmentation criteria and the results from our numerical integrations, and summarize our findings and broader implications in Section \ref{sec:discussion}.  Throughout the paper we assume the $\Lambda\text{CDM}$ cosmological parameters:
$\Omega_{\rm m} = 0.27$, $\Omega_{\Lambda} = 0.73$, and  $h=0.70$, consistent with the 5 yr. WMAP results  \citep{KOM09}. 

\section{Cooling of Shocked, Metal-Enriched Gas}
\label{sec:cooling}

A cold accretion stream penetrates deep inside the protogalactic halo until it encounters an accretion shock. Subsequently, the shocked gas layer grows in size while cooling isobarically \citep[][]{SK87, CB03}, until it becomes gravitationally unstable and fragments into Jeans-mass-sized clumps.  Our objective is to calculate the initial masses of the fragments and  we do not follow their evolution past the point of initial fragmentation. To this end, we model the evolution of the shocked, cooling gas accurately within the one-zone approximation. 
We utilize a comprehensive chemical network to track the time evolution of the abundances of 
all primordial species (H, H$^{+}$, H$^{-}$, $\text{H}_2$, $\text{H}_2^{+}$, e$^{-}$,  He, $\textrm{He}^+$, D, D$^{+}$,
and HD) and the most important metal coolants expected
at high redshifts: neutral and singly ionized O, Si, C, and Fe \citep{BL03, SS06, MD07, GJ07}.  
The cooling gas is subject to compression by the ram pressure of the accretion shock which suggests isobaric conditions are approximately valid in the post-shock region.
In Sections \ref{sec:cooling_processes} and \ref{sec:heating_processes} we proceed to discuss cooling and heating processes in the shocked gas. In Section \ref{sec:chemical_processes} we discuss the relevant chemical processes in the post-shock gas. Finally, in Section \ref{sec:isobaric_cooling} we describe the time evolution of the shock-compressed gas under isobaric conditions.  

\subsection{Cooling Processes}
\label{sec:cooling_processes}

We calculate all relevant cooling and heating processes in low-density, shock heated gas at high redshift. The cooling of 
gas with primordial abundances set by big bang nucleosynthesis has been extensively
studied \citep[e.g.,][]{SK87, OH02}. Above
$10^{4} \, \text{K}$, hydrogen and helium excitation are the
dominant cooling channels. Between $200\,\text{K} \lesssim T \lesssim 10^{4}\,
\text{K} $, if molecular hydrogen forms, the cooling that it facilitates is dominant. If present, HD can continue to cool the gas below $200\,\text{K}$ due to its permanent
dipole moment and smaller rotational energy level spacing, although an elevated electron fraction is necessary for its efficient formation \citep{UI00,NU02,JB06,Y07}. Metals, supplied from previous star formation, can be a significant coolant at $T < 10^{4} \,\text{K}$ due to forbidden, fine structure photon emission
from low-lying electronic radiative transitions \citep{SD93, GS07} and can substantially change the gas dynamics \citep{OMU00,BL03}.
Below $10^{4}\,\text{K}$, our cooling rate as a function of
metallicity closely matches the cooling rates calculated by other authors in gas of similar temperatures and densities
\citep[e.g.,][]{SS06,MD07, SMI08}. We consider fine structure emission from O, Si, Fe, and C---the species that dominate metal cooling in a low density, low metallicity gas \citep[e.g.,][]{SS06}. We find that in the shocked gas in an atomic cooling halo, Si, Fe, and C are, to a large extent, singly ionized (see Section \ref{sec:radiation_field}), but O remains mostly neutral. Above $10^{4}\,\text{K}$, our cooling rate is not accurate at high metallicities
(especially for $Z \gtrsim Z_{\odot}$) as we do not include the multiply ionized species, but since we restrict ourselves to $Z <
10^{-2}\,Z_{\odot}$ (see Section \ref{sec:initial_conditions}), this assumption does not affect our results \citep[see Figure 4 in][]{GS07}.

Due to NLTE conditions when simulating galactic accretion from the IGM,
Boltzmann statistics does not describe the state of an atom or molecule, but instead,
level populations have to be calculated from the principle of
detailed balance \citep[see][]{GJ07}, where electronic
collisional excitations and de-excitations are balanced by spontaneous
downward transitions. Excitation and de-excitation rates, energy level spacings, and spontaneous emission coefficients for C, $\text{C}^{+}$ Si, $\text{Si}^{+}$, Fe, $\text{Fe}^{+}$, O, and $\text{O}^{+}$ are the parameters needed to determine the fine structure cooling rate. We refer the reader to other sources, such as \citet{HM89} (hereafter HM89), \citet{SS06}, and \citet{GJ07}, for detailed discussion of the individual cooling rates. We here only provide the references from which we compiled our data: C -- HM89 and \citet{JOH87}; $\text{C}^{+}$ -- HM89 and \citet{WB02}; O -- \citet{GJ07} and \citet{BEL98}; Si -- HM89; $\text{Si}^{+}$ -- HM89; \citet{ROU90}; Fe, $\text{Fe}^{+}$ -- HM89.

The simplest model of fine structure cooling from an atom is one in which the electron has only one excited state to occupy, the so-called two level system. In this system, the excited state level population is given by \citep[e.g.,][]{MD07}
\begin{equation}
n_{2} = \frac{ \gamma_{12}^{\text{H}} + \gamma_{12}^{\text{e}}n_{\text{e}}/n_{\text{H}} }   {
\gamma_{12}^{\text{H}} + \gamma_{21}^{\text{H}} +
(\gamma_{12}^{\text{e}} + \gamma_{21}^{\text{e}})n_{\text{e}}/n_{\text{H}}  + A_{21}/n_{\text{H}}
}n_{\text{tot}}\mbox{\ ,}
\label{excited_state_level_Pop.}
\end{equation}
where $n_{2}$ is the number density of particles in the second
excited state, $n_{\text{tot}}$ the total number density of the
species, while $\gamma_{12}^{\{\text{H},e\}}$ and $\gamma_{21}^{\{\text{H},e\}}$ are, respectively, the 
excitation and de-excitation rates due to collisions with neutral hydrogen and free electrons (collisions with hydrogen ions, $\textrm{H}_2$, and other species are subdominant). The corresponding volumetric cooling rate for fine structure line emission is given by $\Lambda_{\text{line}} = n_{2}A_{21}\Delta E_{21}$, where $A_{21}$ is the spontaneous emission coefficient and $\Delta
E_{21}$ is the energy difference between states. For $n$ level systems, $n$ coupled equations are solved to determine the
level populations. The CMB temperature sets a thermodynamic lower limit on the gas temperature \citep[see][]{LAR98};
we incorporate this limit by replacing the cooling rates with the effective rate: $\Lambda_{\text{eff}}(T) = \Lambda (T) - \Lambda(T_{\text{CMB}})$.

\begin{figure}
\begin{center}
\includegraphics[width=9.0cm,height=6.5cm]{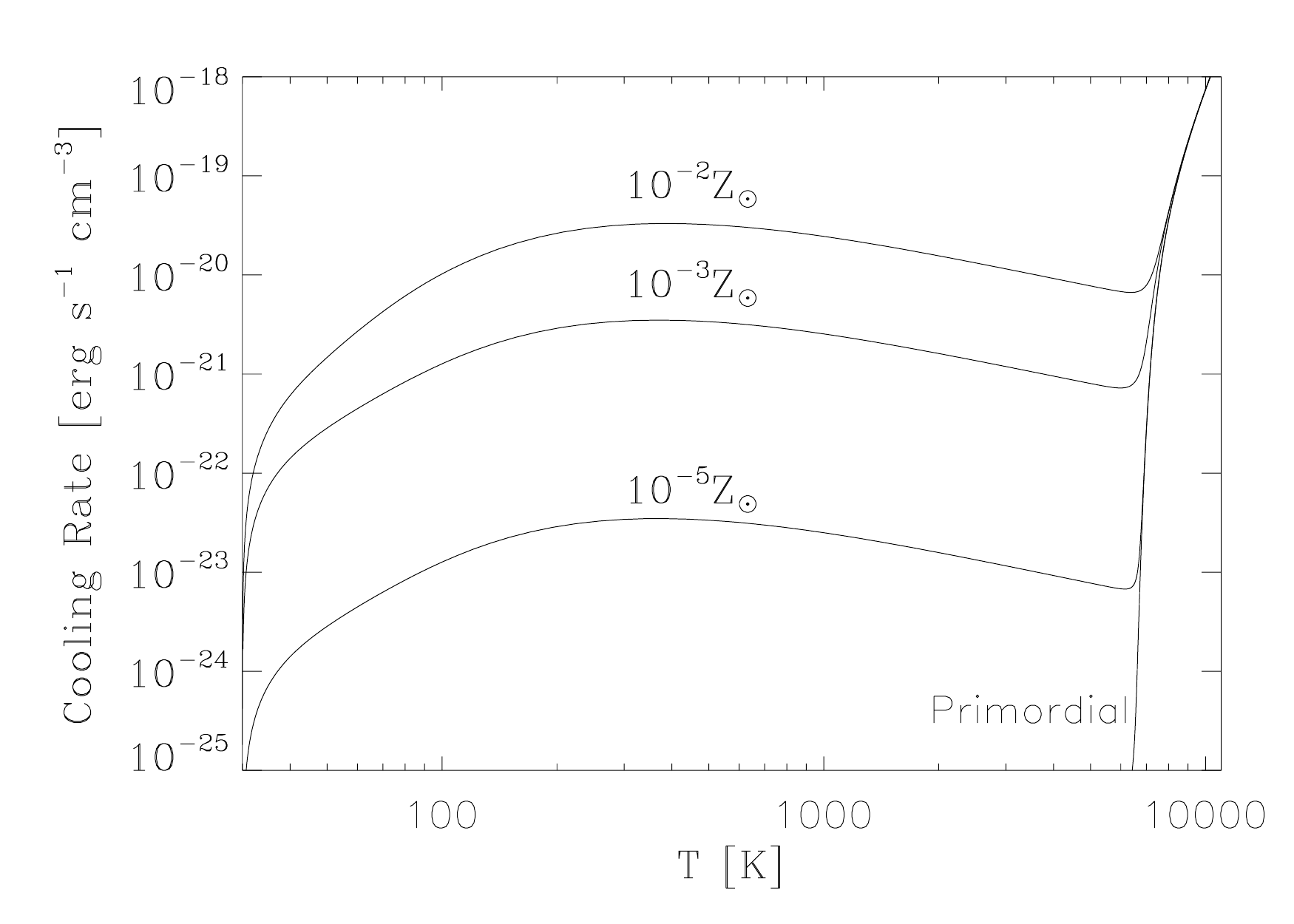}
\caption{Post-shock non-equilibrium cooling rate for differing metallicites for a $10^{8}\, M_\odot$ at $z=10$, using a solar pattern of metals. This figure
only includes cooling from hydrogen excitation and metallic fine structure emission - molecules are ignored. The bump at $T\sim200\,\text{K}$ is due to the
density increase towards lower temperatures due to our isobaric density evolution. The initial density
is $n  \sim 4\times10^{3} \,\text{cm}^{-3}$. Towards $30\, \text{K}$ the cooling rate drops off as we
approach $\text{T}_{\text{CMB}}$. Below $T\sim10^{4}\,\text{K}$ the cooling rate is a strong function of metallicity.}
\label{cooling_rate_metallicity}
\end{center}
\end{figure}

\begin{figure}
\begin{center}
\includegraphics[width=9.0cm,height=6.5cm]{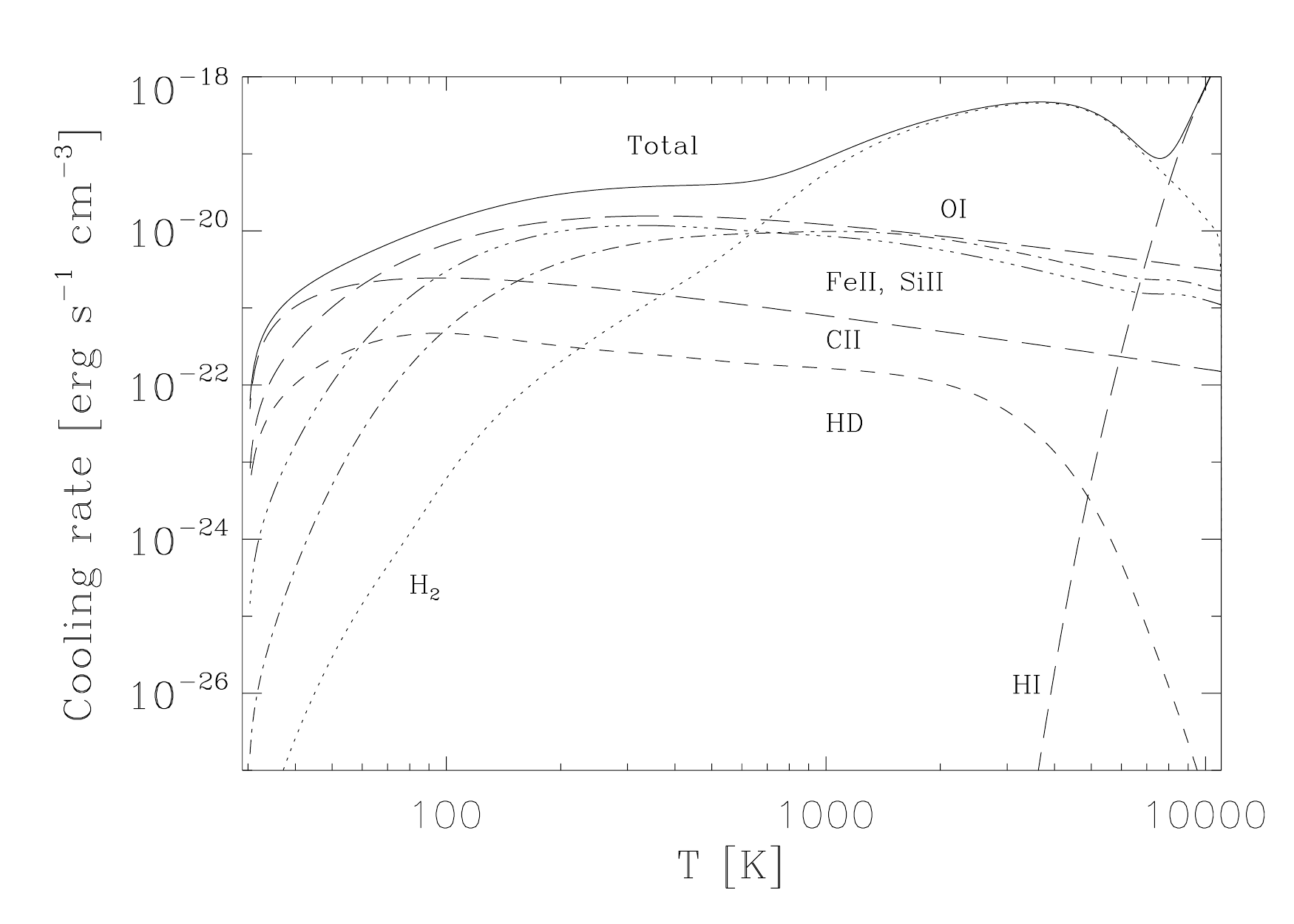}
\caption{Post-shock non-equilibrium cooling rate showing the contributing components for an accretion shock in a $10^{8}\, M_\odot$ DM halo at $z=10$ with $J_{21} = 10$ and metallicity $Z = 10^{-2}\,Z_{\odot}$ with a solar pattern of metals. It is evident that fine structure emission from $\text{Fe}^{+}$, $\text{Si}^{+}$ and O are the dominant metal coolants. $\text{C}^{+}$ is not effective as the density is above its critical density, $n_{\text{cr}, \text{C}^{+}} \simeq 3\times 10^{3}\, \text{cm}^{-3} $.
 \htwo cooling dominates from $1000 - 6000\, \text{K}$ while metals dominate below 1000 K. HD is always an insignificant coolant at $10^{-2}\,Z_{\odot}$.}
\label{noneq_cooling_molecules}
\end{center}
\end{figure}

For illustration, in Figure~\ref{cooling_rate_metallicity} we show the non-equilibrium cooling rate as a function of metallicity, evaluated at every step during the evolution of the post-shock gas in our fiducial model with a $10^{8}\, M_\odot$ halo at $z=10$, solar distributions of metals, and with a high LW radiation background ($J_{21, \text{LW}} = 10^{4}$, see Section \ref{sec:radiation_field}). The latter significantly photodissociates H$_2$, and will keep metallic species singly ionized (see Section \ref{sec:radiation_field}). The wide peak at $T\sim200\,\text{K}$ is due to the isobaric increase in density towards lower temperatures. The initial values for density and temperature are
$4\times10^{3} \,\text{cm}^{-3}$ and $1.1\times10^{4}\, \text{K}$, respectively, and the initial chemical abundances are set to the values corresponding to thermal and statistical equilibrium at this temperature. Below $T\sim10^{4}\,\text{K}$, the cooling rate is a strong function of metallicity.
Figure~\ref{noneq_cooling_molecules} shows the individual elemental contributions to the nonequilibrium cooling curve for a solar abundance pattern. Metal cooling, particularly from O, $\text{Fe}^{+}$, and $\text{Si}^{+}$, dominates below $800\,\text{K}$; $\textrm{H}_2$ dominates between $800$ and  $8,000\,\text{K}$ .The critical density for the $\text{C}^{+}$ fine structure transition is $n_{\text{cr,$\text{C}^+$}} = A_{10} / \gamma_{10} \simeq 3\times10^{3}\,\text{cm}^{-3}$. Thus, at the densities sampled by our isobarically cooling gas, $n = 10^{3} - 10^{6}\,\text{cm}^{-3}$, the $\text{C}^{+}$ line is in LTE, which makes it a weaker coolant compared to $\text{Fe}^{+}$, $\text{Si}^{+}$, and O, whose critical densities are $\sim 10^{5}-10^{6}\,\text{cm}^{-3}$, so thermalization occurs at a later evolutionary stage \citep{SS06}. In this regime, HD is always an ineffective coolant. If, however, metallicity is $Z < 10^{-5}\,Z_{\odot}$, and molecules are permitted to form, the gas is able to reach $T_{\text{CMB}}$ by HD cooling.

\subsection{Heating Processes}
\label{sec:heating_processes}

Since we here ignore the effect of dust, the predominant source of heating is UV pumping of $\textrm{H}_2$, particularly at higher densities. To compute the rate, we assume a pumping rate $8.5$ times that of the photodissociation of $\textrm{H}_2$ \citep{DB96, GJ07}. If each pumping produces $\sim2 \,(1 + n_{\text{cr}} / n)^{-1}\,\text{eV}$ of heat \citep{BUR90}, where $n_{\text{cr}}$ is the critical density for $\textrm{H}_2$ (which we always take to be $6,000\, \text{cm}^{-3}$), the corresponding heating rate is: $\Gamma = 2.7\times 10^{-11}R_{\text{diss}}n_{\htwoeq}(1 + n_{\text{cr}} / n)^{-1} \,\text{erg}\, \text{s}^{-1} \,\text{cm}^{-3}$; the computation of $R_{\text{diss}}$ is discussed in Section \ref{sec:radiation_field} below. 

Other significant sources of heating are the photoionization of metallic species and the radiative coupling to the CMB. It is safe to assume that at $z\sim 10$, prior to complete reionization, the mean free path of H ionizing photons is very short compared to photons below the Lyman limit, thus rendering heating from H and He ionization unimportant. However, photons below the Lyman limit should be $1/f_{\text{esc}} \sim 10^{2}$ times more abundant than H-ionizing ones \citep{BL031}, making photoionization heating of metallic species with first ionization potentials below 13.6 eV (C, Si, and Fe) dominant. This also has consequences for the photodissociation of $\textrm{H}_2$, which occurs in the LW bands just below 13.6 eV (see Section \ref{sec:radiation_field}). As discussed in Section \ref{sec:cooling_processes}, CMB heating is achieved by modifying the effective cooling rate. Overall, heating of the gas is generally not a significant effect for the evolution considered here.

\subsection{Chemistry}

\label{sec:chemical_processes}

{Our chemical model includes all relevant primordial chemical species: H, H$^{+}$, H$^{-}$, $\textrm{H}_2$, $\text{H}_2^{+}$, e$^{-}$, He, $\textrm{He}^+$, $\text{He}^{++}$, D, D$^{+}$, and HD \citep{HAI96, GP02}. We also include the most important metal coolants---neutral and singly ionized C, Si, O, and Fe. Collisional ionization, recombination, and charge exchange reactions for C, Si, and O are taken from the compilation in \citet{JAP09}. The reactions for Fe were taken from \citet{HM89} and \citet{WOOD81}.
We reproduce the complete catalog of rates in our chemical network in Appendix A, using the same numbering convention as in \citet{JAP09} for rates 1-222; after 222, we start our own numbering convention. We initialize our chemical model with equilibrium abundances at $T=1.1\times10^{4}\,\text{K}$. These initial abundances are $n_{\text{H}} \simeq 4 \times 10^{3}\, \text{cm}^{-3}$, $n_{\text{H}^{+}} / n_{\text{H}} \simeq 35$, $n_{\text{H}^{-}} / n_{\text{H}} \simeq 3\times10^{-5}$, $n_{\text{H}_{2}} / n_{\text{H}} \simeq 10^{-2}$, $n_{\text{H}_{2}^{+}} / n_{\text{H}} \simeq 10^{-5}$, $n_{\text{e}} / n_{\text{H}} \simeq n_{\text{H}^{+}} / n_{\text{H}} \simeq 30$, $n_{\text{HD}} / n_{\text{H}} \simeq 10^{-6}$, $n_{\text{He}^{+}} / n_{\text{H}} \simeq 8\times10^{-3}$, and $n_{\text{He}^{++}} / n_{\text{H}} \simeq 4\times10^{-9}$. Metal species are initialized such that Si, Fe, and C are fully ionized while O has the same ionization fraction as hydrogen.

Chemically, the formation and destruction of $\textrm{H}_2$/HD has the strongest impact on our results as it can be a very effective low-temperature coolant. In the absence of dust, molecular hydrogen can form through two channels \citep[e.g.,][]{AB97, TEG97}. In the $\text{H}^-$ channel, $\text{H}^-$ and H combine to form $\textrm{H}_2$  and $\text{e}^-$, whereas in the $\htwoeq^{+}$ channel, $\htwoeq^+$ and H participate in a charge exchange reaction producing $\textrm{H}_2$ and $\text{H}^+$. The efficiency of these formation channels strongly depends on the abundance of free electrons that are required for $\text{H}^-$ and $\htwoeq^+$ synthesis \citep{OH02}. In all our calculations, the $\text{H}^-$ channel dominates. Destruction of $\textrm{H}_2$ is mainly from UV photons in the LW bands and will be discussed in Section \ref{sec:radiation_field}.
HD may be an important coolant below $200\text{K}$ \citep{JB06}. Its formation proceeds by  an exchange reaction between $\text{D}^+$ and $\textrm{H}_2$. In our physical scenario, HD will only be a significant coolant in a low metallicity environment.

\subsection{Isobaric Cooling and Time Evolution}
\label{sec:isobaric_cooling}

If the post-shock cooling is thermally stable and if the shocked gas is in approximate pressure balance, the thermodynamic evolution of a gas packet that has passed the shock obeys conservation of energy, as expressed by \citep[see, e.g.,][]{SK87}:
\begin{equation}
\frac{d(\rho\epsilon)}{dt } = \left(\epsilon + \frac{p}{\rho} \right)\frac{d\rho}{dt} - \Lambda + \Gamma ,
\label{eq:energy_conservation_3}
\end{equation}
where $\rho$, $\epsilon$, $p$, $\Lambda$, and $\Gamma$,  are, respectively, the gas density, specific internal energy, pressure, volumetric cooling and heating  rate. To express these in terms of the chemical species number densities $n_i$, adiabatic indices $\gamma_i$, and the temperature $T$, we write: $p = \sum_{i} n_{i} k_{\text{B}} T$ and $\rho\epsilon = \sum_{i} n_{i}k_{\text{B}}T/(\gamma_{i} - 1)$.
The density derivative $d\ln\rho/dt$ can be written as the total number density derivative $d\ln n/dt$, where  $n = \sum_{i} n_{i}$, minus the number density derivative due to pure chemistry evolution (isochoric), as follows
\begin{eqnarray}
 \frac{d\,\ln\,\rho}{dt} &=&\left(\frac{d\,\ln\,n}{dt} \right)_{\text{fixed}}      \nonumber \\
   &=& \frac{d\,\ln\,n}{dt} - \left(\frac{d\,\ln\,n}{dt} \right)_{\text{isochoric}} \nonumber \\
   &=&-\frac{d\,\ln\,T}{dt} - \left(\frac{d\,\ln\,n}{dt} \right)_{\text{isochoric}}
\end{eqnarray}
where in the last step, we imposed isobaricity, $d\ln\,n/dt = -d\ln\,T/dt$. 'Fixed' refers to evolution with fixed abundances.  With this, equation (\ref{eq:energy_conservation_3}) becomes

  \begin{eqnarray}
 \frac{d}{dt}\left(\sum_{i} \frac{n_{i}k_{\text{B}} T}{\gamma_{i} - 1} \right)  &=&  - \sum_{i} \frac{\gamma_{i}n_{i}k_{\text{B}} }{\gamma_{i} - 1}  \nonumber \\
   && {}\times \left[\frac{dT}{dt} + \frac{T}{n}\left(\frac{dn}{dt} \right)_{\text{isochoric}} \right] \nonumber \\
   && {} -\Lambda + \Gamma
   \label{eq:evolution}
\end{eqnarray}
  
We further calculate the change of species abundances as
\begin{equation}
\label{eq:species_change}
\frac{d\,\ln n_i}{dt} = \frac{d\,\ln\,\rho}{dt}  +  \left(\frac{d\,\ln\,n_i}{dt} \right)_{\text{isochoric}} \ \ \ \ (i = 1,...,N_{\text{species}}) .
\end{equation}
Equation (\ref{eq:evolution}) and (\ref{eq:species_change}) are solved for $dT/dt$ and $dn_i/dt$, and the resulting $N_{\text{species}}+1$ differential equations are integrated implicitly using the Bulirsch-Stoer-type, semi-implicit
extrapolation mid-point method \citep[see][]{BD83}.  The coarsest Bulirsch-Stoer time step is set to $\Delta t = \,\text{min}\{n_{i} / |\dot{n_{i}}|\}$.

Isobaricity, i.e., pressure equilibrium, is a valid approximation if the sound crossing
time of the post shock region is shorter than the cooling time, $t_{\text{sound}} < t_{\text{cool}} $. In Figure~\ref{isobar}, we confirm that this condition is satisfied throughout the evolution of the gas leading to gravitational instability and fragmentation, except for the highest metallicity case, $Z=10^{-2}\,Z_\odot$, when the cooling time becomes shorter than the sound crossing time at $\sim2\times10^{4}\,\text{cm}^{-3}$. We thus conclude that for high metallicity, thermal instabilities may play a dynamically significant role. It is interesting that these instabilities only occur in the presence of efficient metal cooling. Thermal instabilities \citep[e.g.,][]{FIE65,BAL86} are one proposed mechanism for the formation of molecular clouds and the origin of turbulence. \citet{NIK09} performed two-dimensional adapative mesh refinement (AMR) simulations of colliding flows with non-zero metallicity and found turbulent, clumpy structures formed from thermal instabilities triggered by a supersonic collision. It is tempting to suggest that the thermal instabilities we anticipate here at substantial metallicities could lead to clumpiness and turbulence like that seen in \citet{NIK09}.
 
 \begin{figure}
\begin{center}
\includegraphics[width=9.0cm,height=6.5cm]{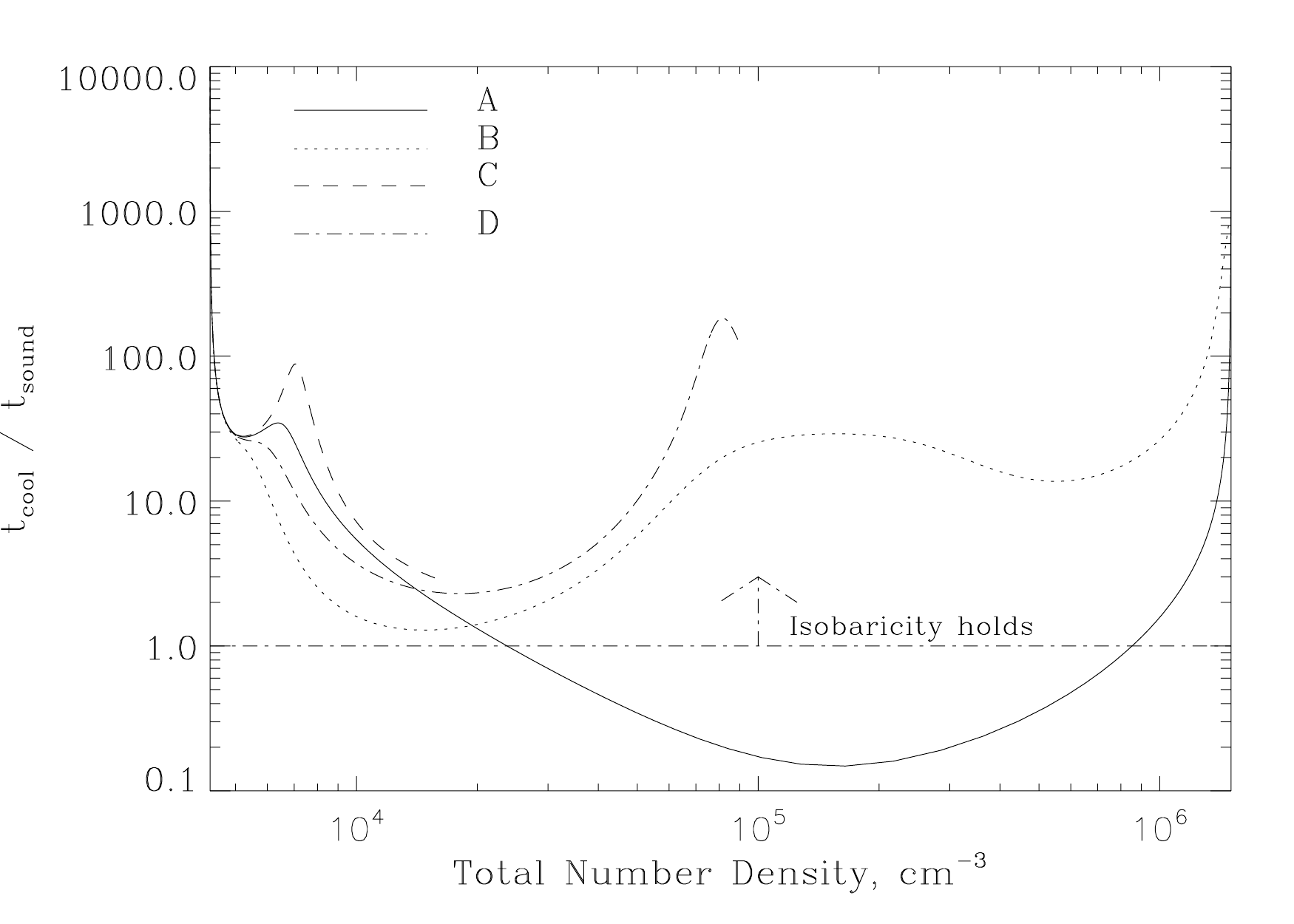}
\caption{Test of the validity of our isobaric density
evolution assumption. The four lines correspond as shown to the four cases of Table \ref{tab:parameters}. As shown, isobaricity holds except in case A (strong metal cooling), where it is violated for densities above $\sim 10^{4}$\,cm$^{-3}$. This will likely lead to runaway cooling collapse in the density regime where $t_{\text{cool}} / t_{\text{sound}} < 1$. In the other cases, where \htwo cooling dominates, we conclude that pressure support
acts on a smaller timescale than radiative cooling. This effectively allows
the ram pressure of the cold flow to keep the shocked region at
approximately constant pressure throughout the evolution, at least until fragmentation occurs. Runs C and D undergo fragmentation before being able to reach high densities, and we do not follow their evolution past this point.}
\label{isobar}
\end{center}
\end{figure}
 
\section{Model Parameters}

\label{sec:initial_conditions}

The choice of parameters for our modeling of the accretion flow into the protogalaxy will govern the resulting fragmentation mass scale. This section focuses on the three most important model ingredients. In Section \ref{sec:cold_flow_accretion}, we discuss the properties of the cold flows into high redshift galaxies. In Section \ref{sec:abundances_early_enrichment}, we describe the possible abundance patterns of accreting gas and place an upper limit on the metallicity content of the gas. Finally, in Section \ref{sec:radiation_field}, we discuss the effect of a LW radiation background on the dissociation of molecular hydrogen and the ionization of metallic species.

\subsection{Cold Flow Accretion}
\label{sec:cold_flow_accretion}

\citet{BD03} consider accretion of non-virialized gas from the IGM into a DM halo and find that, at redshifts $z>2$,  under the assumptions of spherical symmetry, a stable virial accretion shock does not exist in halos smaller than $10^{11}\,M_\odot$ due to the rapid cooling of the post-shock gas.  Our model halos are well within this regime.  Instead, the accreting gas continues on supersonic infall trajectories until it reaches much closer to the center of the halo.   \citet{DB06} argue that such `cold streams' provide a natural explanation for many observed galactic phenomena, including the existence of starbursts lacking the evidence of a major merger, the existence of bulgeless disks, and the blue-red bimodality observed in present-day galaxies.  \citet{KER05} find, using smoothed particle hydrodynamics (SPH) simulations, that in halos with masses $< 10^{11.4}\,M_\odot$ at $z=3$, almost none of the gas that enters the galaxy has been heated to the virial temperature, further supporting the conclusion that cold accretion is the main feeder of baryonic material into galaxies at high redshift. 

\citet{WA07} simulated the cosmological formation of a $\sim 10^8\,M_\odot$ cosmological object and find that the virialization of shocked accretion streams creates a nearly isothermal ($\sim r^{-12/5}$) density profile containing supersonically turbulent gas.
\citet{GB08} carried out a simulation a similar mass halo at $z\sim 10$ and found that 
filamentary accretion dominates the buildup of the baryonic component of the galaxy. 
In their simulation, the density of cold accretion streams just prior to impact with the central turbulent mass concentration was $\sim 10^{3} \,\text{cm}^{-3}$ with a temperature of $\sim 200\,\text{K}$. 
The pre-shock number density in cold streams can be estimated by considering infall in a halo with mass $M=10^8\,M_8\,M_\odot$ and virial radius $R_{\rm vir}=1.5 \,M_8^{1/3}\,[(1+z)/10]^{-1}\,\text{kpc}$, where the latter we define as the radius within which the mean density of a halo equals $200$ times the critical density of the universe. If the density of the infalling baryons scales as $\propto r^{-2}$, the hydrogen number density extrapolated to the innermost $10~\text{pc}$, the approximate radius of cold-stream collisions and shock formation \citep{Wise:08a}, is $\sim 10^3~\text{cm}^{-3}$, consistent with the numerical result.

\subsection{Abundances from Early Enrichment}
\label{sec:abundances_early_enrichment}

We restrict ourselves to an upper metallicity limit of $10^{-2}\,
Z_{\odot}$. This follows from the work of \citet{GB08} who find that
a $10^{8}\, M_\odot$ halo will be assembled out of material that has experienced a maximum of $\sim10$
Pop III SNe. \citet{BYH03} showed that in a Pop III SN explosion, $\sim 90\%$ of the
 metallic yield will be ejected into the IGM. \citet{GREIF10} find that accretion inflows return roughly 50\% of the metallic enrichment to the central
halo by the time it grows to $10^{8}\, M_\odot$. Assuming a homogeneous density
 distribution of metals over the halo, roughly 1 proper kpc in size, gives a metallicity fraction of $\sim 10^{-3}
\,Z_{\odot}$, suggesting that $10^{-2}\,Z_{\odot}$ is a
conservative upper limit for the metallicity. We explore the effect of metallicity that varies from
a primordial composition, which is effectively described by $Z\leq 10^{-8}\, Z_{\odot}$, to this upper limit, using both PISN and solar abundance patterns.

There have been assertions that a LW background from Pop III stars
could inhibit further star formation by destroying $\text{H}_{2}$ and $\text{H}^-$, which would keep the first galaxies starless and thus metal free or metal deficient \citep[e.g.,][]{HB06, Dijk08}. Other
studies have shown that an ionizing background enhances metal
enrichment as a high rate of ionization encourages molecule
formation \citep[e.g.,][]{JOH08} and enhanced cooling. Another recent study
revisited the question of radiative feedback from Pop~ III
stars using detailed radiation-hydrodynamical simulations and found radiative
feedback to be neutral \citep{AS07} for Pop III.2 star formation (stars still forming from pure H/He gas but with radiative feedback from previous star formation), suggesting there will
be a metal abundance of at least $10^{-4} \,Z_{\odot}$ in
these protogalaxies, depending on the halo capture fraction of
SN nucleosynthetic products. The metal abundance pattern is also not known with any certainty. \citet{UN02, UN03} suggest that the abundance pattern reflects enrichment from hypernovae, produced by stars in the $20-120\,M_\odot$ mass range, while other authors argue that Pop~III stars end their lives as PISNe, producing a very different metallicity pattern \citep{HW02}. Therefore, it is crucial to explore a range of abundance patterns in our modeling.

Furthermore, the mixing of heavy elements
within the protogalaxy after they get incorporated through cold accretion streams, or following a SN explosion inside the galaxy itself, remains poorly understood. Heavy elements produced from Pop~III stars may never return to the galaxy and may reside in the IGM indefinitely. Another possibility is that metals may leave the host galaxy after a SN and return at a later time. Once the galaxy is assembled, will there be a region of high metallicity contained around SNe, or will metallicity quickly diffuse throughout the protogalaxy? Inhomogeneous mixing may create pockets of enhanced enrichment, with a metallicity much greater than $Z_{\odot}$. The degree of mixing or clumpiness of the metal distribution inside the protogalaxy will be investigated in greater detail in later work.

Our numerical integrations are run for both standard solar \citep[see][]{GS98}  and PISN metal abundance patterns. PISN patterns are from \citet{HW02} for
a 150 and $250\, M_\odot$ progenitor SN (corresponding to a 75 and $125 \,M_\odot$
helium core, respectively). Table \ref{tab:abundance_pattern} represents the number density (in $\text{cm}^{-3}$) of metallic species given a hydrogen number density of $1\, \text{cm}^{-3}$. The PISN values are computed with the same overall mass fraction in metals, but differing elemental proportions compared to the Sun.

\begin{deluxetable*}{ccccc}
\tablecolumns{5}
\tablewidth{4.2in}
\tablecaption{Metallicity Yield Patterns. These values represent, for a gas with a hydrogen number density of $1\, \text{cm}^{-3}$, the number density of metallic species wtih differing abundance patterns normalized to the metallic mass fraction of the sun.
\label{tab:abundance_pattern}}
\tablehead{\colhead{Pattern} & \colhead{C} & \colhead{Fe} & \colhead{O}  & \colhead{Si}  }
\startdata
Solar & $ 2.5\times10^{-4} $ & $2.8\times10^{-5}$ & $4.6\times10^{-4}$  & $3.2\times10^{-5}$ \\
$150 \,\text{M}_{\odot}$ PISN & $5.0\times10^{-5}$ & $1.2\times10^{-6}$ & $5.1\times10^{-4}$ & $1.4\times10^{-4}$  \\
$250 \,\text{M}_{\odot}$ PISN & $1.4\times10^{-5}$ & $1.3\times10^{-4}$ & $1.5\times10^{-4}$ & $1.0\times10^{-4}$
\enddata
\end{deluxetable*}

\subsection{Radiation Field}
\label{sec:radiation_field}

The rate of photoionization of a species ($R_{\text{photo}}$) for an optically thin gas is given by \citep[e.g.,][]{OST89}
\begin{equation}
R_{\text{photo}} = \int_{\nu_{\text{th}}}^{\infty} \frac{4\pi J_{\nu}}{h\nu}\sigma(\nu)    \,d\nu ,
\label{eq:photoionization_rate}
\end{equation}
where $J_{\nu} = 10^{-21}\,J_{21}\,\text{ergs}\, \text{s}^{-1} \,\text{cm}^{-2}\, \text{Hz}^{-1}\, \text{sr}^{-1}$ is the mean intensity, $\sigma(\nu)$ is the frequency dependent photoionization cross section, and the integration runs from the threshold photoionization frequency, $\nu_{\text{th}}$, to infinity. In the case of metallic photoionizations which are dominant (see Section \ref{sec:heating_processes}), we run this integration to $\nu_{\text{H}}$, the ionization threshold of hydrogen, as hydrogen ionizing photons are effectively shielded by the IGM prior to reionization. Photoionization cross-sections for C, Si, Fe, and O are taken from the fits of \citet{VER96}. For our incident spectrum, we use thermal Planck spectra with $T_{*}=10^{4} $ and $10^{5}\,\text{K}$ and a power law spectrum, $J_{\nu} \propto \nu^{-1}$. As in \citet{BL03}, we normalize these spectra at the Lyman limit by $J_{21}$.
The photodissociation of $\textrm{H}_2$ is a more complicated process. It occurs through the two-step Solomon process \citep{SW67} by photons in the LW bands, $11.2 - 13.6 \,\text{eV}$. A commonly used approximation for this rate of dissociation, $R_{\text{diss}, \htwoeq}$, is given by $\sim 1.38 \times 10^{9} J_{\bar{\nu}}$ where $J_{\bar{\nu}}$ is the mean intensity at 12.4 eV, the mean energy of a LW band photon \citep[see][]{AB97}. We assume the photodissociation rate of HD equals that of $\textrm{H}_2$.

At high column densities, $\textrm{H}_2$ can be particularly effective at shielding itself from LW radiation which can significantly alter its photodissociation rate \citep[e.g.,][]{OH02}. \citet{DB96} provide an approximation for the shielding factor such that $R_{\text{diss}} \propto f_{\text{shield}}$, where $f_{\text{shield}} \simeq \text{min}[1.0, (N_{\htwoeq}/10^{14}\text{cm}^{-2})^{-0.75}]$ with $N_{\htwoeq}$ being the column density of $\textrm{H}_2$. However, this applies to a static, cold medium. If the LW bands are Doppler shifted by large thermal or bulk motions, the shielding rate will be much lower. In in accreting protogalaxy, where large-scale velocity gradients reach $\sim 20 \;\text{km} \;\text{s}^{-1}$, self-shielding will certainly be overestimated within this prescription. Taking thermal broadening into consideration, \citet{DB96} provide another fit to the shielding factor as
\begin{equation}
 f_{\text{shield}} = \frac{0.965}{(1 + x/b_{5})^{2}} + \frac{0.035}{(1+x)^{0.5} }\, \text{exp}[-8.5\times10^{-4}(1+x)^{0.5} ] ,
\end{equation}
 where 
$x = N_{\htwoeq}$ and  $b_{5} = b / 10^{5}\; \text{cm}\; \text{s}^{-1}$.
We follow \citet{AS07} in setting $b = 9.12\; \text{km}\; \text{s}^{-1}$, as their physical system is approximately similar to ours. Although this will still overestimate the amount of self-shielding when bulk motions dominate, it is an improvement over the previous fit and will be used in computing our value of $\textrm{H}_2$ self-shielding. 

The radiation field present in these high redshift protogalaxies is uncertain, owing to the contribution from the external, global IGM field, and the local field produced by internal Pop III/II stars. One can estimate, at $z=10$ , that the intensity just above the Lyman limit, $J^{+}_{21}$, necessary to reionize the universe is $\sim 40$ \citep{BL03}. However, photons with energies below the Lyman limit are virtually free to escape their host halo, so that the corresponding intensity, $J^{-}_{21}$, which includes the $\textrm{H}_2$ dissociating LW bands, could potentially be $f_{\text{esc}}^{-1} \sim 100$ times higher, where $f_{\rm esc}$ is the escape fraction of H-ionizing photons \citep{WL00}. Other recent estimates  \citep{RS09,Wise:09} have found $f_{\text{esc}} \sim 0.25-1$ in dwarf galaxies at high redshift. Finally, the LW background can be elevated orders of magnitude above the global average LW field in clustered halos \citep{Dijk08, WHA08,JOH08, AH09, SHA09} . For this study, we take $J_{21} = 10$ and $J_{21} = 10^{4}$, representing small and large radiation fields which could be realistically present at this epoch.
 
\section{Thermal Evolution and Fragmentation}

\label{sec:thermal_evolution}

We now consider the case of a cold accretion stream which is shock heated in the center of a primordial protogalaxy. In Section \ref{sec:frag_criteria} we derive a simple criterion for when the post-shock region will fragment based on the free-fall and sound crossing times. Section \ref{sec:results} describes the thermal evolution of the post-shock region, and presents the fragmentation mass scale for different parameter choices, whereas Section \ref{sec:toy_model} outlines a simple analytical model for the critical metallicity in molecule-free regime.

\subsection{Fragmentation Criterion and Mass Scale}
\label{sec:frag_criteria}

To derive a rough estimate of when fragmentation will occur, we consider the growth of the shocked gas region within our one-zone model.
Following the shock that terminates the accretion stream near the center of the protogalaxy, the gas begins to cool off isobarically \citep[see][]{SK87, CB03} from the virial temperature of the halo, $T_{\text{vir}} =  \onehalf \mu m_p v_{\text{vir}}^{2} / k_{\text{B}} \simeq 2\times 10^{4} \mu_{0.6} M_8^{2/3}[(1+z)/10] \,\text{K}$, where $\mu=0.6\,\mu_{0.6}$ is the mean molecular weight in the immediate post-shock gas.
Applying the standard Jeans analysis for gravitational instability, we take it that fragmentation occurs when the ever-increasing sound crossing time equals the decreasing free fall time of the post-shock region.
Specifically, we examine three time scales: the cooling
time, $t_{\text{cool}} \simeq\frac{3}{2}k_{\text{B}}Tn_{\text{tot}}/\Lambda(n, T)$, the free-fall time, $t_{\rm ff} = (3\pi/32G\rho)^{1/2}$, and the sound-crossing time, $t_{\rm s}
\simeq L/c_{\rm s}$, where $L$ is the length of the shocked slab, $\Lambda(n,T)$ the volumetric cooling rate, and $c_{\text{s}} = (\gamma k_{\text{B}} T/\mu m_p)^{1/2}$ the sound speed. The pre-shock density $\rho_0$ is related to the instantaneous post-shock density $\rho_1$ via $\rho_{0}(v_{\text{infall}} + v_{\text{shock}}) = \rho_1v_{\text{shock}}$. Since the shock is strong, $\rho_1\sim 4\rho_0$. If the post shock gas further cools and isobarically compresses to $\rho(t)$, then the shock velocity in the frame of the gas will be further reduced by a factor of $\sim \rho_1/\rho(t)$, or $v_{\rm shock}/v_{\rm infall}\sim \rho_0\rho_1/[\rho(t)(\rho_1-\rho_0)]\sim \frac{4}{3}\rho_0/\rho(t)$, and thus the length of the post shock region is (ignoring numerical factors order unity),
 \begin{equation}
L(t) \sim v_{\text{shock}} t \sim \frac{\rho_0}{\rho(t)}\: v_{\text{vir}} \, t ,
\label{eq:region_size}
\end{equation}
where $v_{\text{infall}} = v_{\text{vir}} = (GM/R_{\rm vir})^{1/2}\simeq 23\, M_8^{1/3} [(1+z)/10]^{1/2}\,\text{km s}^{-1}$ the virial (or infall) velocity.
Isobaric cooling requires $t_{\rm cool} > t_{\rm s}$, while gravitational (Jeans) instability and the onset of fragmentation requires  $t_{\rm s} > t_{\text{ff}}$. 
It is possible that gravitational instability does not set in before the gas cools to the CMB temperature.  In this case, isobaricity is guaranteed and the fragmentation mass scale is entirely set by the CMB temperature and the initial conditions of the accretion flow.

We assume that following the initial gravitational instability, the gas fragments into clumps with masses given by the local Bonnor-Ebert (BE) mass, 
\begin{equation}
\label{eq:BE}
M_{\text{BE}} \simeq 150\,M_{\odot}\,(T/200\,\text{K})^{2},
\end{equation}
 assuming isobaric density evolution with $T_{i} = T_{\text{vir}} = 1.1\times10^{4}\,\text{K}$ and $n_{i} = 4.0\times10^{3}\,\text{cm}^{-3}$.
The resulting clump size sets an upper limit to the masses of the star in the aftermath of the fragmentation. Due to subtle processes affecting post-fragmentation accretion, the final stellar masses could be a fraction of the fragmentation scale: $M_\star \simeq \alpha M_{\text{BE}}$, where $\alpha < 0.5$ in star formation in the local universe \citep{MT02}.

\subsection{Case Studies}
\label{sec:results}

\begin{deluxetable}{clllc}
\tablecolumns{5}
\tablewidth{3in}
\tablecaption{Parameters for the Different Runs
\label{tab:parameters}}
\tablehead{\colhead{Run} & \colhead{Metallicity} & \colhead{$J_{21}$} & \colhead{$T_{*}$} & \colhead{$M_{\text{BE}}$ Final}}
\startdata
A......... & $ 10^{-2}\, Z_{\odot} $ &          $10^{4}$        &        $10^{4}$ &   $\sim3\,M_{\odot}$ \\
B......... & $10^{-6}\, Z_{\odot}$ &            $10      $         &        $10^{4}$ &   $\sim10\,M_{\odot}$ \\
C......... & $10^{-6}\, Z_{\odot}$ &           $10^{3}$         &        $10^{4}$ &   $\sim10^{5}\,M_{\odot}$ \\
D......... & $10^{-6}\, Z_{\odot}$ &           $10^{3}$         &        $10^{5}$ &   $\sim10^{4}\,M_{\odot}$ 
\enddata
\end{deluxetable}

For our case studies, we consider a high and low metallicity case ($10^{-2}$ and  $10^{-6}\, Z_{\odot}$), a high and low LW radiation fields ($J_{21} = 10, 10^{3}$, and $10^{4}$), and source temperatures of $T_{\ast} = 10^{4}$ and $10^{5}\,\text{K}$.  We do not present a run with high metallicity and efficient molecular cooling, but we do comment on the outcome of such a run at the end of this section. In Table \ref{tab:parameters} we provide the parameters of our four representative runs. We present the results for each case in succession. See Figures~\ref{fig:temp_time}--\ref{fig:ion_degree} for the temperature history, cooling rate, $\textrm{H}_2$ fraction, and ionization degree (defined such that $\text{ionization degree} \equiv n_{e^{-}\text{free}} / n_{p^{+}\text{any form}} $), respectively, for each run..

Run A represents a relatively high metallicity environment with a strong LW background that photodissociates $\textrm{H}_2$, rendering metals the predominant coolant at $T < 10^{4}\,\text{K}$. As in all our runs, we assume that the baryons in DM filaments penetrate deeply into the center of a protogalaxy where they undergo a strong shock with ${\cal M} = v_{\text{vir}} / c_{\text{s}} \sim 10$. Calibrating to  the gas density in \citet[][]{GB08} as explained in Section \ref{sec:cold_flow_accretion} above, the post-shock number density is $4n_{0} \sim4\times10^{3} \,\text{cm}^{-3}$. The shock raises the gas temperature to $T_{\text{vir}} \sim 1.1\times 10^{4}\,\text{K}$. At this temperature, the gas cools effectively via atomic hydrogen line excitation, while the size of the post-shock region begins to increase in size according to equation (\ref{eq:region_size}). After $\sim10^{5}\, \text{yr}$ the gas has cooled to $\sim35 \,\textrm{K}$, which is slightly above $T_{\text{CMB}}(z=10) \simeq 30\,\text{K}$ due to UV pumping of $\textrm{H}_2$, and after $\sim1.5 \times 10^{6}\, \text{yr}$ the post-shock region has reached the fragmentation criterion ($t_{\text{ff}} < t_{\text{sound}}$). We do not follow the evolution past this point. The radiation background ($J_{21} = 10^{4}$) has prevented an appreciable amount of $\textrm{H}_2$ from building up, with $f_{\htwoeq} \equiv n_{\htwoeq} / n_{\text{H}} $ never exceeding $10^{-7}$. The BE mass at the point of fragmentation is $\sim 3 M_{\odot}$. Using a fit to the baryonic growth rate of halos of $\dot{M} \simeq 6.6\, M_{12}^{1.15}(1+z)^{2.25}f_{.165}\,M_{\odot} \,\text{yr}^{-1}$ \citep{DEK09}, where $M_{12} = M_{\text{vir}} / 10^{12}\,M_{\odot}$ and $f_{.165}$ is the halo baryonic fraction in units of the cosmological value $\Omega_{\rm b}/\Omega_{\rm m}\simeq 0.165$, we find that the protogalaxy accreted $\sim 5\times 10^{4}\, M_{\odot}$ of baryons before the onset of fragmentation. If all these baryons experienced fragmentation, this would represent a region with $\sim 10^{4}$ separate fragments, possibly the precursor to a stellar or globular cluster \citep[see][]{BC02}. For more realistic star formation efficiencies, of order $\sim 0.1$, we would still
expect a sizable cluster of Pop II stars to arise.

Run B explores a low metallicity environment together with a feeble LW background which allows for $\textrm{H}_2$/HD formation and cooling. Starting from the same initial density and temperature as Run A, molecular hydrogen, now able to form more effectively, reaches an abundance $f_{\htwoeq} \sim 10^{-3}$ only $10^{3}\, \text{yr}$ after the shock passage; $\textrm{H}_2$ and HD dominate the cooling from $8,000$ to $30 \,\text{K}$ (see Fig.~\ref{fig:cooling_rate_by_run}). Like in Run~A, the post-shock region fragments after $\sim 10^{6}\, \text{yr}$ at a temperature of $\sim50\,\text{K}$. The final BE mass is $\sim 10\,M_{\odot}$. Overall, this run is very similar to case A, but cooling is now dominated by $\textrm{H}_2$ and HD instead of metallic fine structure lines. The likely end product is a collection $\sim 10^{3-4}$ fragments.

Run C is intended to illustrate the result for a post-shock region with little metal enrichment and a large enough LW background to keep $\textrm{H}_2$ cooling suppressed, similar to the case examined in \citet{BL031}. Initially, the gas is still able to cool via atomic hydrogen excitation to $\sim 6,000\,\text{K}$, when the cooling time becomes similar to the Hubble time. The gas thus stalls at $6,000\,\text{K}$, at a density of $6\times10^{3} \,\text{cm}^{-3}$, until it fragments $10^{6}\, \text{yr}$ after the shock. The final BE mass is $10^{4}\,\text{M}_{\odot}$ while the total accreted mass is also of this order. This accretion episode, with no effective low temperature coolant, produced one large fragment, possibly indicative of an object which will not fragment further and become a super massive black hole (SMBH) precursor \citep{OMU01,BL03, SHA09,Schleicher:10}. Because of our overestimation of $\textrm{H}_2$ self-shielding (see Section \ref{sec:radiation_field}), this value reflects a lower limit on the fragmentation mass scale, which is most likely higher, possibly large enough to facilitate one large $10^{5}\,M_{\odot}$ fragment capable of direct collapse to a SMBH.

Run D has the same initial conditions as Run C, except that the blackbody source temperature is now $T_{*} = 10^{5}\,\text{K}$ instead of $10^{4}\,\text{K}$, more representative of the type of sources associated with Pop III stars \citep{BK01}. Because of the normalization at 13.6 eV, a blackbody of $T_{*} = 10^{5}\,\text{K}$ is much less effective at destroying $\text{H}^{-}$, so that $\textrm{H}_2$ formation is inhibited much less than for a $T = 10^{4}\,\text{K}$ blackbody (see Fig.~\ref{fig:h2_frac}). The slightly more effective $\textrm{H}_2$ cooling leads to a final fragmentation mass scale of $10^{3}\,M_{\odot}$. Like in case C, overestimation of $\textrm{H}_2$ self shielding sets this as a lower limit on the fragmentation mass scale.
 
  \begin{figure}
\begin{center}
\includegraphics[width=9.0cm,height=6.5cm]{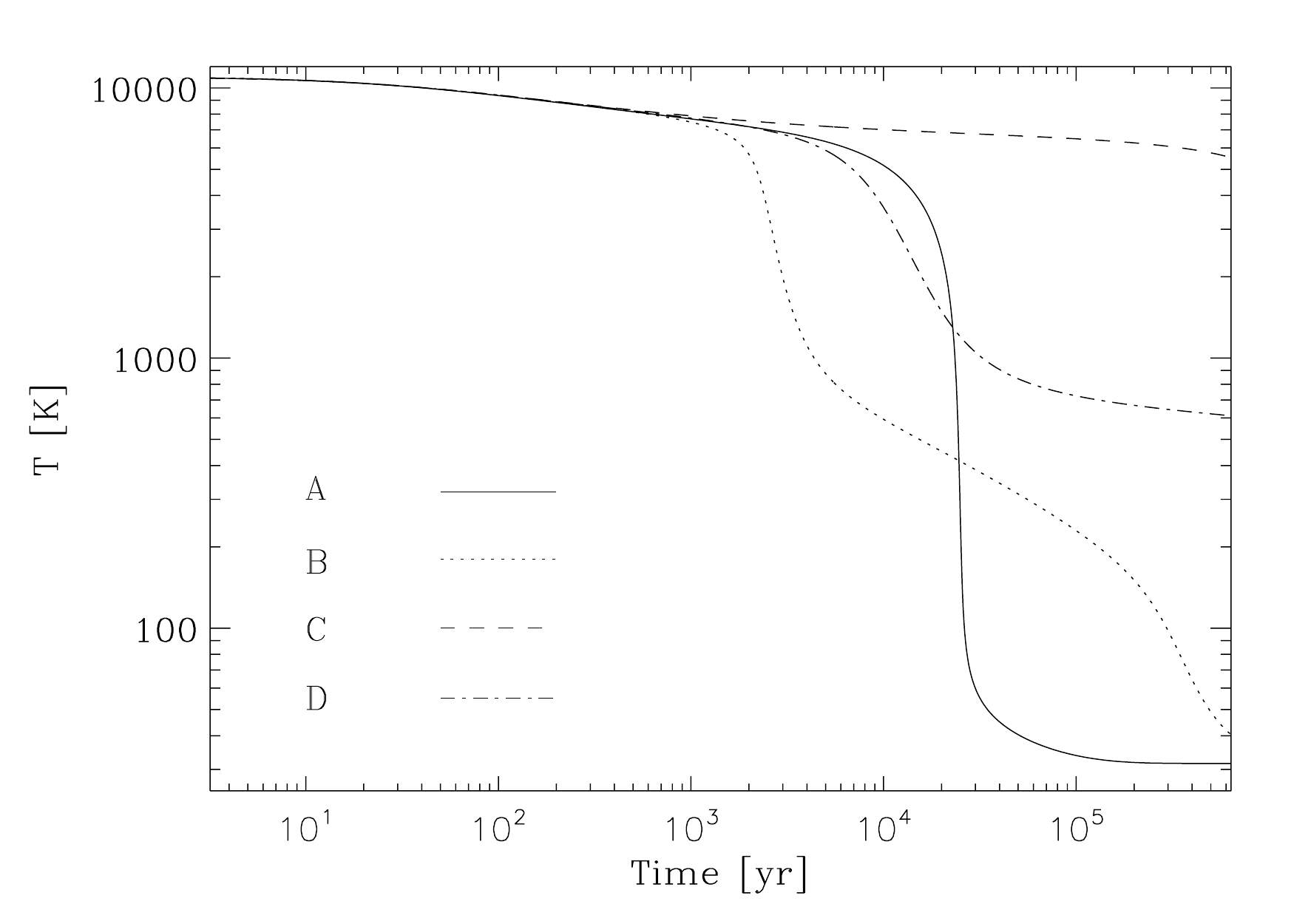}
\caption{Post-shock temperature history of four different runs, all performed with a solar abundance pattern. See Table \ref{tab:parameters} for a description of the run parameters. All runs start at $T = 1.1\times10^{4}\,\text{K} \sim T_{\text{vir}}$. The addition of molecules (\htwo predominantly) produces additional cooling at high temperatures ($T\sim 10^{3}\, \text{K}$) and thus the temperature falls off from $T \sim 6500\,\text{K}$ first. Metals, which are effective low temperature coolants, allow the gas to reach the $T_{\text{CMB}}$ quickly, as shown in case A. Case C initially has no effective coolant below $T \sim 10^{4}\,\text{K}$, but as \htwo self-shielding builds up the gas is able to cool a small amount before fragmentation. Case D, with a blackbody source temperature of $T=10^{5}\,\text{K}$, is able to build up a small \htwo fraction and cool to $\sim 500\,\text{K}$ before becoming gravitationally unstable. Fragmentation occurs for each run at $\sim 10^{6} \,\text{yr}$.}
\label{fig:temp_time}
\end{center}
\end{figure}
 
 \begin{figure}
\begin{center}
\includegraphics[width=9.0cm,height=6.5cm]{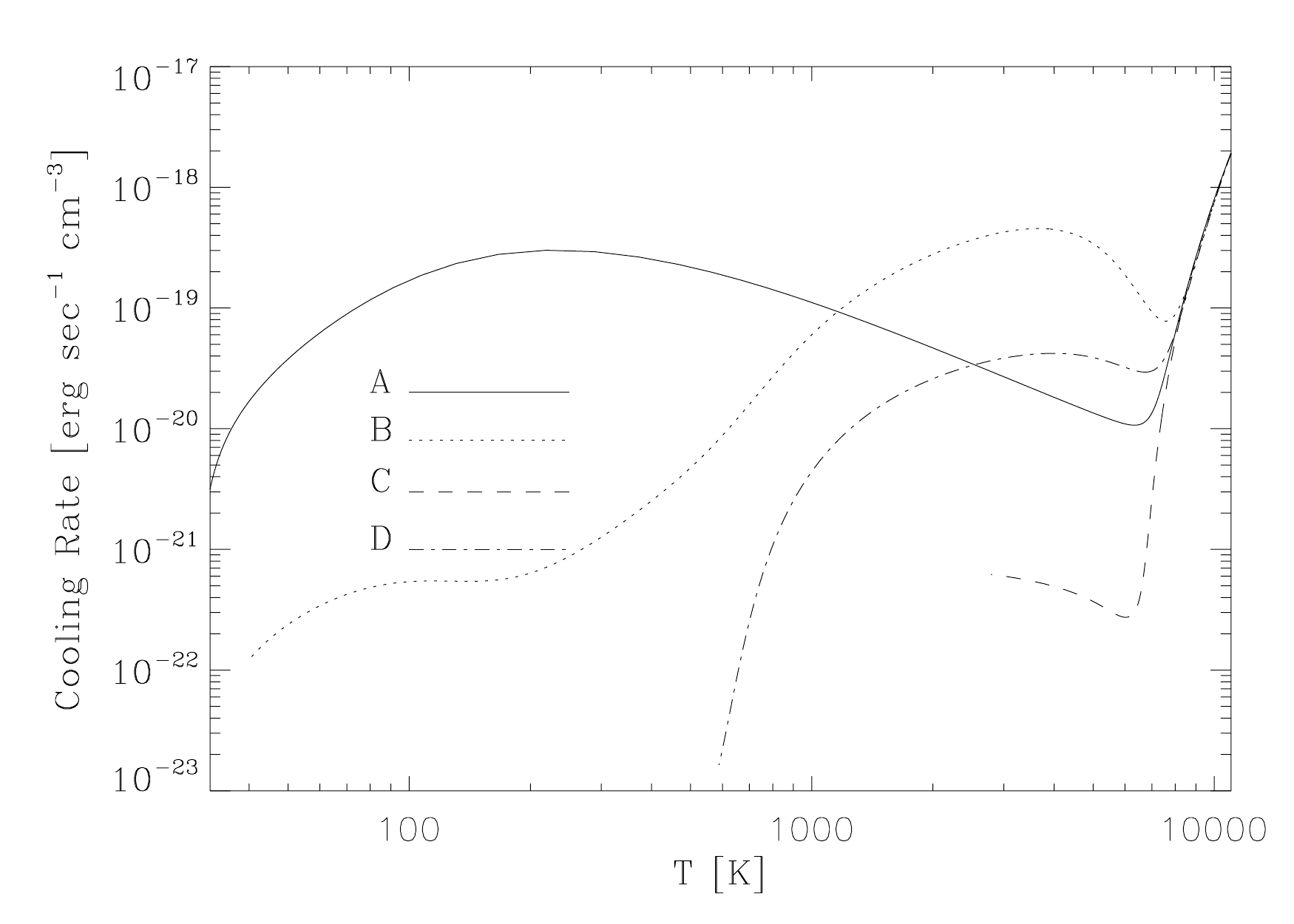}
\caption{Total cooling rate as a function of temperature for the four representative runs. For Cases C and D, which fragment before reaching close to $T_{\text{CMB}}$, only the pre-fragmentation cooling curve is shown. Metals dominate the cooling in case A at $T\sim 200\,\text{K}$ while \htwo is a very effective coolant at $T\sim 3000\,\text{K}$.}
\label{fig:cooling_rate_by_run}
\end{center}
\end{figure}

\begin{figure}
\begin{center}
\includegraphics[width=9.0cm,height=6.5cm]{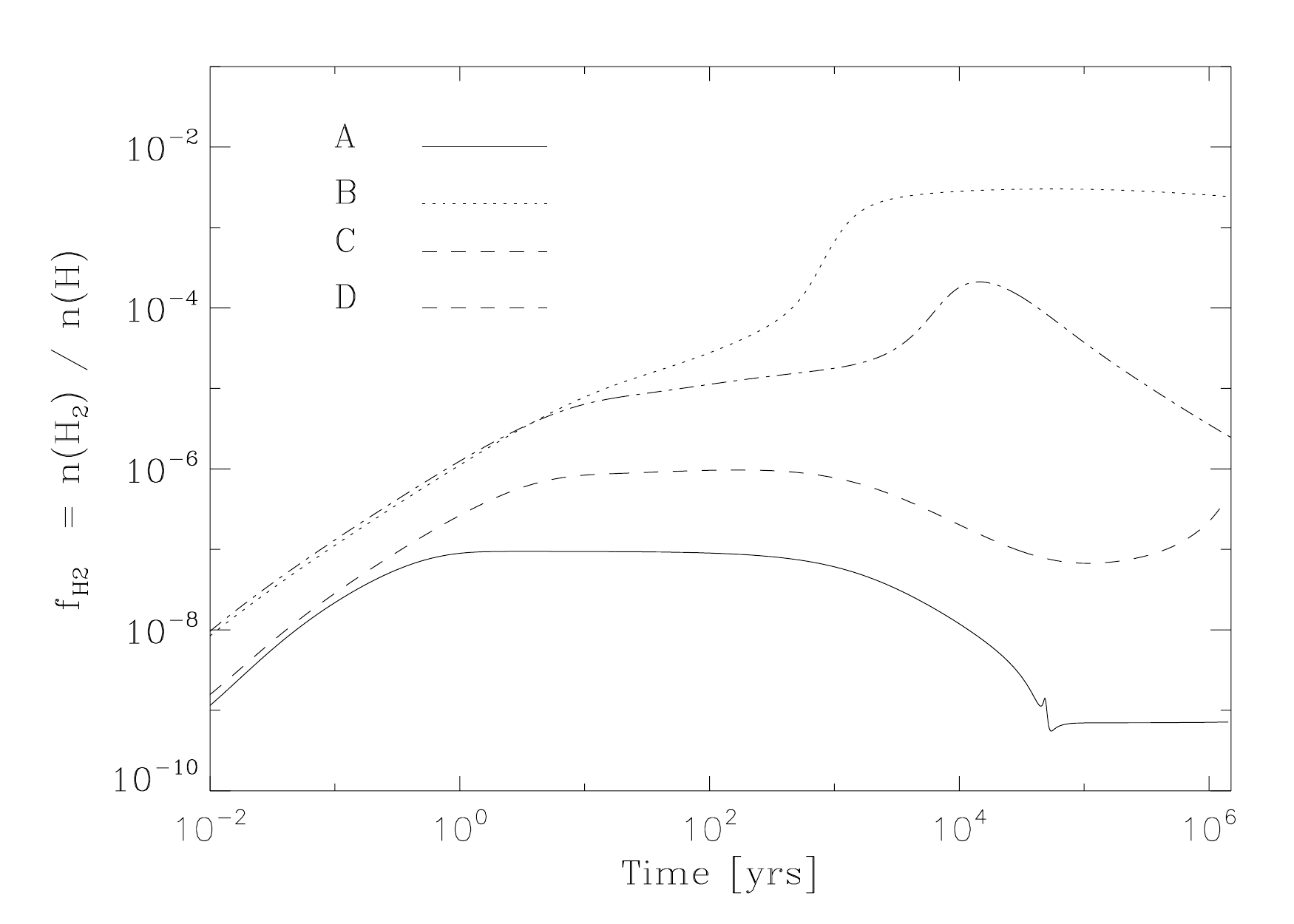}
\caption{Time evolution of \htwo, $f_{\htwoeq}$, in the four representative runs. The strong photobackground in case A prevents any appreciable \htwo fraction from forming, while a LW background of $J_{21} = 10$ allows the \htwo fraction, $n(\htwoeq) / n(\text{H})$, to exceed $10^{-3}$.}
\label{fig:h2_frac}
\end{center}
\end{figure}

\begin{figure}
\begin{center}
\includegraphics[width=9.0cm,height=6.5cm]{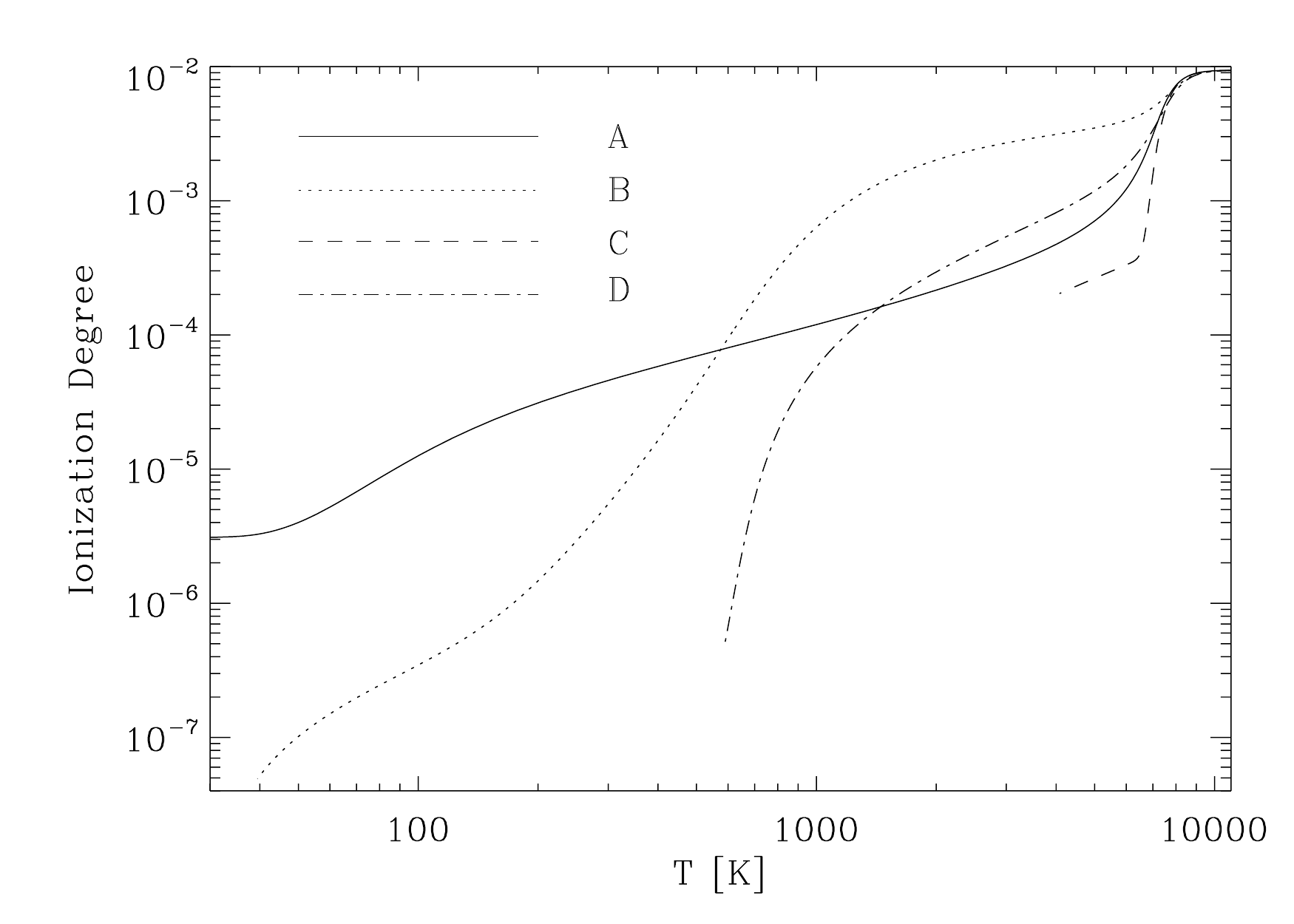}
\caption{Ionization degree of the gas in the four runs. We define the ionization degree to be the number density of free electrons divided by the number density of protons in any form, $n_{e^{-} \text{free}} / n_{p^{+}}$.  }
\label{fig:ion_degree}
\end{center}
\end{figure}

Our results indicate that if there is a low temperature coolant available, either metals or molecules, the gas will fragment with a characteristic mass of $1-10\,M_{\odot}$ (cases A and B); a fragment in this mass range has a very good chance of producing a solar mass star. A run with both efficient molecular cooling and high metallicity ($J_{21} = 10$ and $Z = 10^{-2}~Z_{\odot}$) will behave similarly to case A and produce a $\sim 1~ M_{\odot}$ fragment. If, however, no low-temperature coolant is available as a result of a high LW background and little chemical feedback from early SNe, there will likely be only one large fragment ($\sim 10^{4-5} M_{\odot}$). The critical LW background where $\textrm{H}_2$ is no longer an effective coolant is approximately $J_{21} = 10^{3-4}$ for a blackbody spectrum of $T_* = 10^{4}\,\text{K}$, close to the value found by \citet{BL031} for similar initial conditions. \citet{SHA09} find a much smaller critical $J_{21}$, most likely due to the higher $\textrm{H}_2$ self-shielding experienced in these relatively high density accretion flows, although as discussed, we do consistently overestimate the degree of self-shielding.  \citet{Schleicher:10} find cooling suppression for $J_{21}\sim 1-10^4$, depending on the gas density.
Another point of interest is the appearance of a critical metallicity in the absence of molecular coolants. In case A, the final fragmentation mass is a strong function of metallicity (see Fig.~\ref{frag_mass_scale}). At low metallicities ($Z \lesssim 10^{-4}\,Z_{\odot}$) one large fragment is produced. Conversely, high metallicities ($Z \gtrsim 10^{-4}\, Z_{\odot}$) produce $\sim 10^{3}-10^{4}$ solar mass fragments, lending support to metallicity  being the driver of the Pop III-II star formation transition protogalactic accretion flows. This behavior can be understood by comparing the cooling time from metals and a characteristic fragmentation time, as we explain in Section \ref{sec:toy_model} below. We note that in runs with PISN abundances, a given metallicity will produce fragments nearly one order of magnitude smaller. Yields from PISNe are thus more effective at reducing the fragmentation mass scale. 

 \begin{figure}
\begin{center}
\includegraphics[width=9.0cm,height=6.5cm]{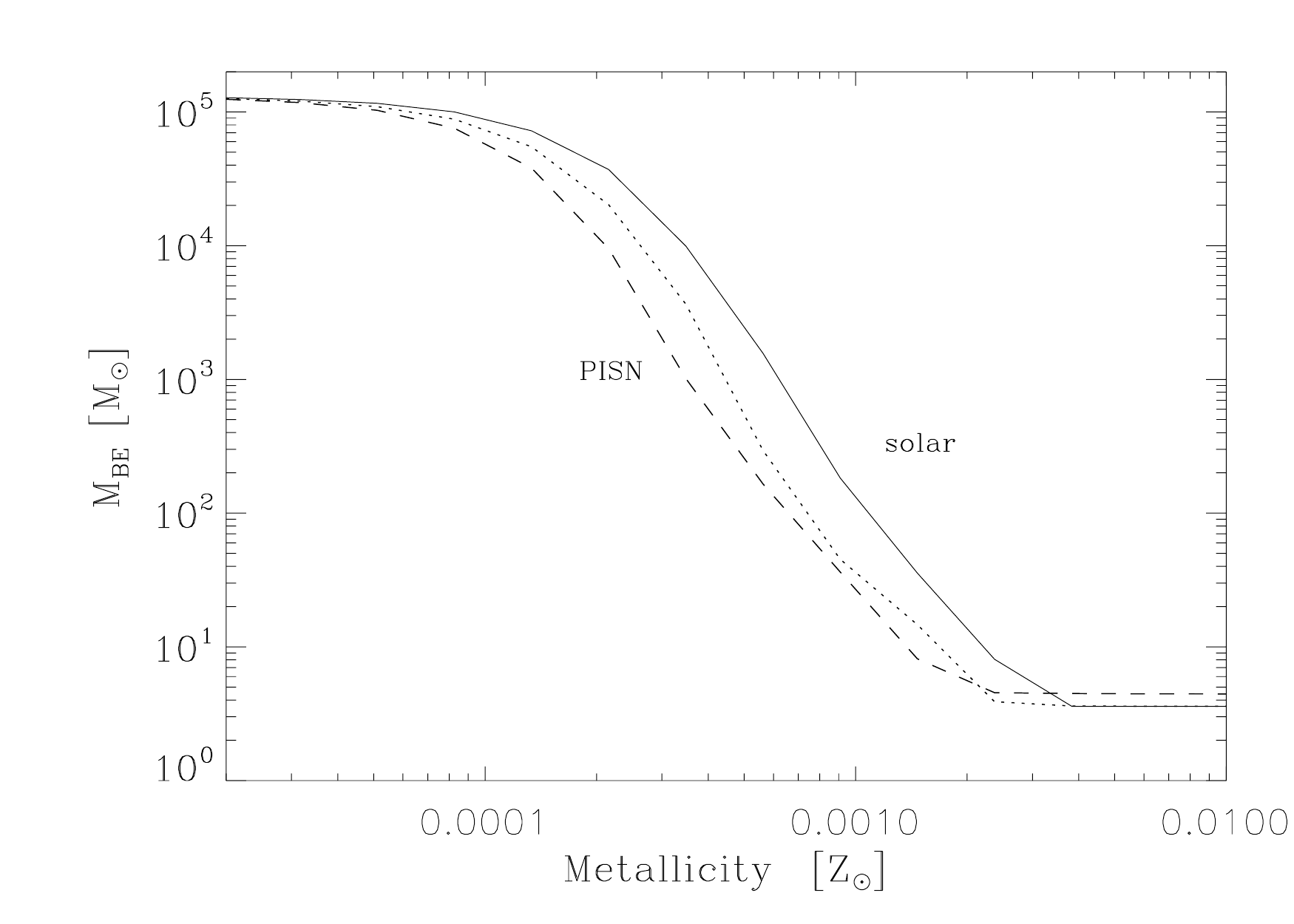}
\caption{Characteristic fragmentation mass as function of
metallicity for three different metallicity abundance patterns. The masses represent the Bonnor-Ebert mass at the point of fragmentation, when $t_{\text{sound}} = t_{\text{ff}}$. These represent the physical scenario in case A ($J_{21} = 10^{4}$) but with varying levels of metallicity. The two curves labeled
PISN (overlaid) are for a 150 and $250\, M_\odot$ PISN metallicity yield pattern.  The critical
metallicity (the metallicity where the fragmentation mass scale drops appreciably)
occurs at slighly lower metallicity for the two runs performed with PISN abundance patterns, due to the higher proportion
of oxygen and silicon produced in PISNe as compared to the solar abundance pattern, the shift is small however.
The upper mass scale of $\sim 10^{5}\, M_\odot$ reflects the initial conditions of the accretion flow while the lower mass limit of $\sim 3 \,M_\odot$ is set by the CMB temperature floor. As discussed in Section \ref{sec:toy_model}, the critical metallicity can be understood by considering the three relevant timescales, $t_{\text{cool}}$, $t_{\text{sound}}$, and $t_{\text{ff}}$; it is dependant
on the gas being able to cool appreciably before the fragmentation criteria is reached (determined by $t_{\text{sound}}$ and $t_{\text{ff}}$).}
\label{frag_mass_scale}
\end{center}
\end{figure}

\subsubsection{The Impact of Dust}
 \label{sec:dust}

The presence of dust, which may accompany metals \citep[e.g.,][]{CAU02}, could have a significant impact on the thermal evolution of the gas in that it radiates thermally and also catalyzes molecular hydrogen formation. Dust grains should be produced in primordial supernovae \citep{TOD01,NOZ03,SCH04}, but uncertainty still exists as to how much dust ultimately enriches the ISM \citep{NOZ07,CHE10}. We do not include dust thermal emission or grain surface $\text{H}_{2}$ and HD formation in our chemical model. Dust thermal emission is always a sub-dominant coolant at metallicities up to $10^{-2}\,Z_{\odot}$ and densities below $10^{6}\,\text{cm}^{-3}$ \citep{OMU05}. To assess the catalytic impact of dust grains in our model, we compare the $\text{H}_{2}$ formation rate on dust grains with the $\text{H}_{2}$ formation rate through the $\text{H}^{-}$ channel and the LW $\text{H}_{2}$ photo-dissociation rate. The formation rate of $\text{H}_{2}$ on dust grains is \citep{CS04}
\begin{eqnarray}
R_{\text{d}} &=& 3.025\times10^{-17}\, \epsilon_{\htwoeq}n_{\text{H}}n(\text{H})\,S_{\text{H}}(T)\,f_{\text{a}}(T)   \nonumber \\
&& \times \, \left(\frac{T_{\text{g}}}{100}\right)^{1/2}\, \left(\frac{\xi_{\text{d}}}{0.01}\right) \,  \text{cm}^{-3}\, \text{s}^{-1},
\label{eq:h2_grain_form}
\end{eqnarray}
where $\epsilon_{\htwoeq}$ is the $\text{H}_{2}$ formation efficiency \citep{CS09}, $S_{\text{H}}(T)$ is the temperature dependent sticking factor of hydrogen onto dust grains \citep{CS04}, $n_{\text{H}}$ is the number density of neutral hydrogen, $n(\text{H})$ is the number density of hydrogen nuclei, and $f_{\text{a}}(T)$ is a temperature dependent factor accounting for thermal evaporation of hydrogen from the grain surface \citep{TH85}. We assume a typical dust grain has a mass of $1.2\times 10^{-14}\,\text{g}$, such that $n_{\text{d}} = 1.3\times 10^{-10}\xi_{d}n_{\text{H}}$. Equation (\ref{eq:h2_grain_form}) is normalized for a standard Galactic dust-to-gas mass ratio of $\xi_{\text{d}} = 0.01$ at around solar metallicity. We can scale this for lower metallicity by writing $R_{\text{d}}(Z) = (Z/Z_{\odot})\,R_{\text{d}}(Z_{\odot})$ where $R_{\text{d}}(Z_{\odot})$ is equation (\ref{eq:h2_grain_form}) evaluated for $\xi_{\text{d}} = 0.01$. 

 \begin{figure}
\begin{center}
\includegraphics[width=9.0cm,height=13.5cm]{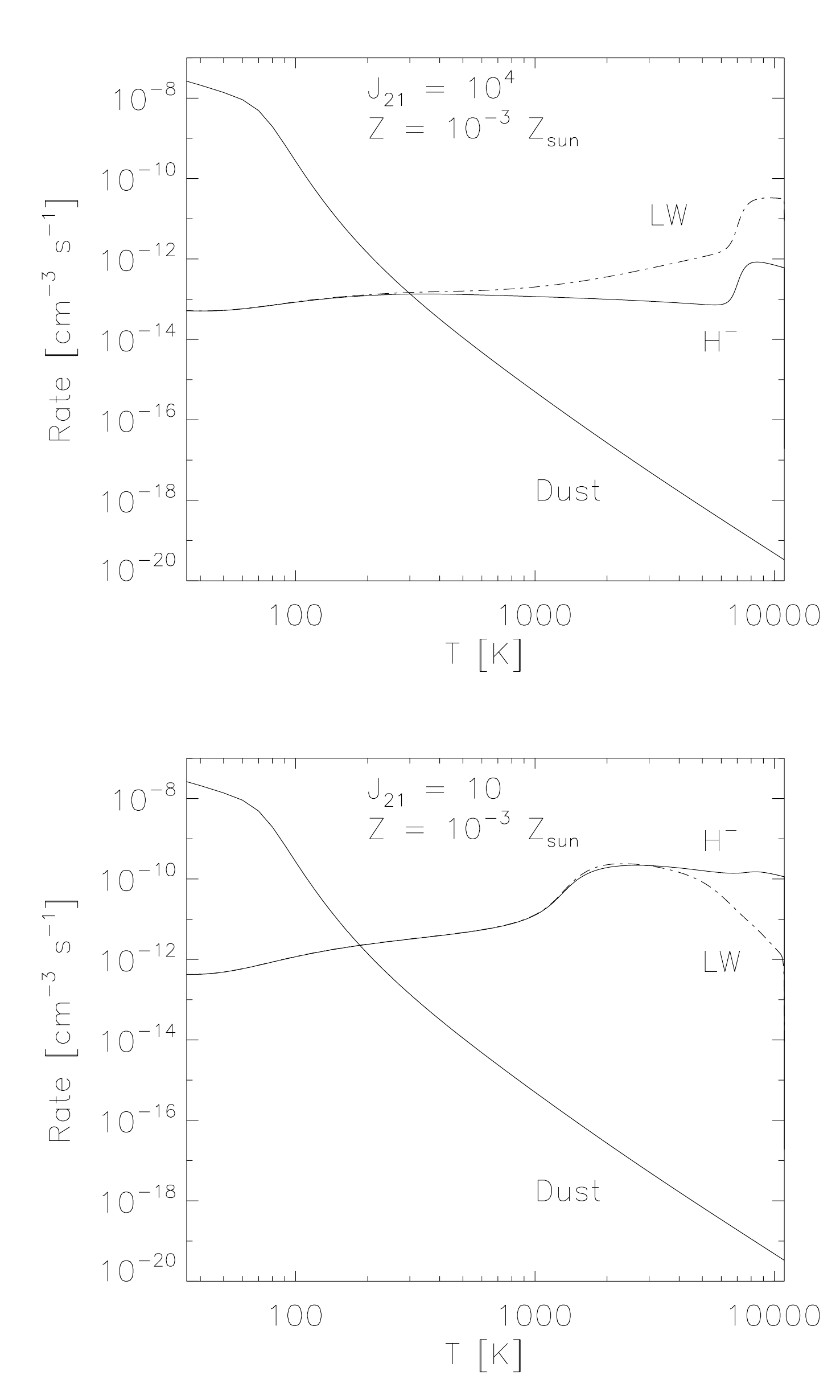}
\caption{The volumetric chemical rates for the LW $\text{H}_{2}$ destruction rate (labeled LW), the formation rate of $\text{H}_{2}$ through the $\text{H}^{-}$ channel ($\text{H}^{-}$), and the grain $\text{H}_{2}$ formation channel rate (Dust) for two different values of the intensity in the LW bands (see Section \ref{sec:radiation_field}); $J_{21} = 10^{4}$ (top) and $J_{21} = 10$ (bottom). The dust formation rate drops off rapidly at high temperatures due to the steep temperature dependences of the H sticking coefficient $S_{\text{H}}(T)$ and the H evaporation coefficient $f_{\text{a}}(T)$. In the high LW intensity run, LW destruction initially dominates the chemistry of $\text{H}_{2}$  while dust formation is negligible. At $600\,\text{K}$, $\text{H}_{2}$ is in equilibrium with the radiation field and at $300\,\text{K}$ dust $\text{H}_{2}$ formation becomes the primary formation channel. Gas which reaches roughly $T\sim250\,\text{K}$ will experience rapid $\text{H}_{2}$/HD formation, allowing the gas to cool rapidly to $T_{\text{CMB}}$.
}
\label{dust_rate}
\end{center}
\end{figure}
In Figure \ref{dust_rate} we compare these three volumetric chemical rates---the LW $\text{H}_{2}$ destruction rate, the $\text{H}_{2}$ formation rate through the $\text{H}^{-}$ channel, and the $\text{H}_{2}$ formation rate on dust grains for two different values of $J_{21}$. The grain $\text{H}_{2}$ formation rate is computed with equation (\ref{eq:h2_grain_form}) using the isobaric density evolution scheme while the other two rates are extracted from our chemical model. In the run with $J_{21}=10^{4}$, LW photon destruction of $\text{H}_{2}$ dominates over dust grain $\text{H}_{2}$ formation at temperatures $T>250\,\text{K}$. Below this temperature, $\text{H}_{2}$ formation on dust grains is extremely efficient. We also repeat this run for $J_{21}=10$ and find similar behavior. We chose these two values of $J_{21}$ as they bracket the critical $J_{21}$ for $\text{H}_{2}$ formation through the $\text{H}^-$ channel (see Section \ref{sec:results}). We do not consider dust opacity and its effect on the LW photon flux, as understanding the interplay between photodissociation and dust requires complicated multidimensional modeling of PDRs \citep[e.g.,][]{KRU08,KRU09}. We also do not consider $\text{H}_{2}$ formation heating in this work, but as is shown in \citet{OMU05}, at metallicities up to $10^{-2}\,Z_{\odot}$ and densities below $10^{6}\,\text{cm}^-3$, this heating mechanism is subdominant  compared to $\text{H}_{2}$, metal line cooling, and adiabatic heating.

 Examining Figure \ref{frag_mass_scale} and using equation (\ref{eq:BE}), $T\sim250\,\text{K}$ corresponds to a BE mass of $\sim230\,M_{\odot}$ which is the initial fragmentation mass scale of a gas with metallicity $Z\sim10^{-3}\,Z_{\odot}$. Molecular hydrogen, however, is not able to cool gas below $T\sim200\,\text{K}$ due to its lack of a permanent dipole moment, thus further cooling depends on the formation rate of HD and other molecules (e.g., CO) which can cool the gas to even lower temperatures---these molecules will most likely also be formed on dust grains in significant number around this temperature as well. We can expect then, because of dust's impact on molecule formation, that the fragmentation mass scale will be sharply lowered at around $10^{-3}\,Z_{\odot}$. At higher metallicity,  $\sim10^{-2}\,Z_{\odot}$, the formation rate of $\text{H}_{2}$ on dust grains will increase. However, gas with a metallicity of $10^{-2}\,Z_{\odot}$ will reach $T_{\text{CMB}}$ through metal cooling alone (see Figure \ref{frag_mass_scale}); the addition of molecular cooling will only decrease the cooling time. Thus, while the formation rate of $\text{H}_{2}$ on dust grains will become increasingly larger at higher metallicity, it will not affect the end product, namely the initial fragmentation mass scale, of the gas in our model. Increased molecule formation brought about by dust may also allow thermal instabilities to develop more easily---see Section \ref{sec:isobaric_cooling}. We conclude that molecule formation on dust grains will alter the behavior of the gas below some critical temperature ($T\sim250\,\text{K}$) making the transition from high mass fragments to low mass fragments occur more abruptly, but while retaining the general picture which we present here.

These findings are complementary to those of \citet{OMU08} who also use a one-zone model to study the thermal evolution of gas in a halo with $T_{\text{vir}} \sim 10^{4}\,\text{K}$. They find, with the inclusion of dust, that the gas will fragment vigorously with a metallicity as low as $Z\sim10^{-6}\,Z_{\odot}$. This episode of dust-facilitated secondary fragmentation occurs at densities $n_{\text{H}}\sim 10^{10}\,\text{cm}^{-3}$, higher than the densities, $n\sim 10^{5-6}\,\text{cm}^{-3}$, at which we first encounter fragmentation. For the densities which our models overlap there is excellent agreement---see Figures 5a and 5b in \citet{OMU08}.

\subsubsection{Initial Density}
 \label{sec:init_density}
  These results depend sensitively on the choice of initial post-shock density, due to the dependence of the cooling rate and the BE mass on density. Motivated by \citet{GB08}, we chose a pre-shock density of $n=10^{3}\,\text{cm}^{-3}$ as this is the density right before the cold stream gas experiences a shock near the center of the galaxy. If the pre-shock number density is smaller, $10^{2}\,\text{cm}^{-3}$, the initial BE mass is $\sim 5\times10^{5}\,M_{\odot}$. The critical metallicity for the gas to reach $T_{\text{CMB}}$ is also correspondingly shifted to higher metallicity, roughly $10^{-2.5}\,Z_{\odot}$, and even at this metallicity, the gas fragments with $M_{\text{BE}}\sim 10\,M_{\odot}$. For a pre-shock density of $10\,\text{cm}^{-3}$, the initial BE mass is $\sim 2\times10^{6}\,M_{\odot}$ and the gas temperature is not able to reach $T_{\text{CMB}}$ at any reasonable metallicity.

\subsubsection{Hot Accretion Mode}
\label{sec:hot_accretion}

We proceed to comment on the implications of an even lower pre-shock density encountered in quasi-spherical baryonic accretion from the IGM onto the halo. As has been shown in simulations \citep{WA07, GB08}, accompanying the cold accretion mode is a hot mode where gas is accreted at the virial radius directly from the IGM residing between the (cold) filaments of the cosmic web. The fate of gas accreted in this mode will differ substantially from the case of cold accretion, mainly in the location of the shock and the subsequent cooling rate. As gas is accreted directly from the IGM, its post-shock density is $\sim0.4\,\text{cm}^{-3}$ \citep{WA07}, much lower than the initial density adopted for this study ($10^{3}\,\text{cm}^{-3}$). Initially, the gas is not self gravitating as indicated by its large BE mass. None of the cooling processes discussed here will allow the gas to cool appreciably below $T\sim7000\,\text{K}$. At this point, $t_{\text{cool}}\sim 10^{9}\,\text{yr}$ while $t_{\text{ff}}\sim3\times10^{7}\,\text{yr}$; this suggests a stable virial shock may develop, even with a high degree of metallicity in the gas. It seems likely that only the gas which enters the galaxy through a dense, cold accretion filament is able to immediately fragment with a small characteristic fragmentation mass.

\subsection{Analytical Toy Model for Critical Metallicity}
\label{sec:toy_model}

As shown in Section \ref{sec:results}, runs with a large LW radiation background ($J_{21} > 10^{4}$) exhibit a strong dependance on metallicity of the final fragment mass scale. Here, we discuss the physical basis behind this critical metallicity, utilizing a simple toy model. As demonstrated, our critical metallicity is a result of cosmological initial conditions via the virial velocity and accretion density, cooling physics, and the Jeans analysis for gravitational fragmentation. 

Let $t_{\text{frag}}$ be the time it 
takes the expanding shocked region to reach the fragmentation criterion:
$t_{\text{sound}}> t_{\text{ff}}$, where $t_{\rm sound}\sim L/c_{\rm s}$, the ratio of the length of the shocked region, given by equation (\ref{eq:region_size}), and the sound crossing time. In the absence of metals and molecular cooling, atomic cooling quickly brings the temperature from its post-shock value of $T\sim 10^4~\text{K}$ to the asymptotic $T \sim 6,500~\text{K}$, resulting in $L_{Z=0}(t)\sim \frac{1}{5}\,v_{\rm vir} t$, where the factor of $1/5$ is from equation (\ref{eq:region_size}) considering adiabatic and isobaric density enhancement at $T=6,500\,\text{K}$.
In metal- and molecule-free gas fragmentation consequently sets in at $t_{{\rm frag},Z=0}\sim 10^6\,\text{yr}$. Interestingly, due to the competing temperature dependence of the free-fall and sound crossing time, $t_{\text{frag}}$ is roughly constant for any value of metallicity. To determine the level of metals needed to allow the gas to fragment at a small mass scale, we evaluate $t_{\text{frag}} \sim t_{\text{cool}}$, where $t_{\text{cool}}$ is the cooling time due to metal transitions. Equating the timescales yields
\begin{equation}
t_{\text{cool,} i}(Z) 
\sim  \frac{k_{\text{B}}T}{X_{i}}A_{10}\Delta
E_{10} \sim 10^{6} \,\text{yrs},
 \label{tfrag_tcool}
\end{equation}
where $A_{10}$ is the spontaneous emission coefficient of a
transition, $\Delta E_{10}$ is the corresponding energy difference
between states, and $X_{\text{i}} \equiv n_{\text{i}} / n_{\text{tot}}$ is the  number density fraction of the metal coolant.  For the solar abundance pattern, O is the dominant coolant over a wide range of temperatures (see
Fig.~\ref{noneq_cooling_molecules}). For the two lower states of O, $A_{10}  = 8.9 \times
10^{-5}\, \text{s}^{-1}$ and $\Delta E_{10} = 3.14 \times 10^{-14}\,\text{ergs}$.  The critical O abundance therefore is 
\begin{equation}
X_{\text{O}} \sim \frac{k_{\text{B}}T
}{10^{6}\,\text{yrs}\,\,A_{10}\Delta E_{10}} \sim 10^{-8} \mbox{\ ,}
\label{critical_oxygen_abundance}
\end{equation}
where $T = 6,500\,\text{K}$. This fraction corresponds to an oxygen mass fraction $Z_{\text{O}} \sim 10^{-4.8} \,Z_{\text{O},\odot}$, agreeing with the critical metallicity demonstrated by Figure \ref{frag_mass_scale}. The same analysis could be performed with the cooling rate of silicon (as it is the most effective coolant for runs with a PISN abundance pattern)---the conclusion would remain the same. Previous studies on the role played by metallicity in the star formation mode transition  \citep{BRO01, OMU05, SS06}  have defined the critical metallicity
as the metal fraction that alters the characteristic mass of stars formed from the standard Pop III initial conditions where cooling by $\textrm{H}_2$ and metals competes with the compressional heating due to self-gravity. Our physical scenario is very different; instead of having a self-gravitating, adiabatic collapse in $10^{6}\, M_\odot$ DM minihalos, we consider an isobarically cooling gas that was shocked to $10^4\,\text{K}$ in $10^{8}\,M_\odot$ halos.  Direct comparison between critical metallicity values obtained in these two contexts is not entirely justified.  Nonetheless, we do find a critical metallicity comparable to some of the minihalo results \citep{BRO01}. It is not clear why a hot, isobarically cooling shock should behave so similarly to the adiabatic collapse experienced in a minihalo. 

\section{Discussion and Conclusions}
\label{sec:discussion}

In this work, we have examined the fate of metal-enriched, cold, dense, and dust-free accretion
streams into DM halos of mass $10^{8}\, M_\odot$ at redshift
$z=10$; we regard these objects as the first galaxies. From simulations, we know these cold streams penetrate very close to the center of the protogalaxy, where they shock and fragment. Our goal was to determine the fragmentation properties of
these cold streams and to identify the key physical parameters governing this first step in the high redshift star 
formation process. We find that metallicity (due to metals being an effective low temperature coolant) and LW background strength (as it is capable of destroying $\text{H}_{2}$ and HD)  are the dominant factors in determining the final fate of the post-shock region. Using a fit to the baryonic growth rate of halos, and assuming, as recent studies suggest, that these galaxies are fed entirely by cold streams and that their star formation rate closely matches the mass accretion rate, we can estimate the number of fragments produced by these flows. We either see one large ($\sim10^{4}M_{\odot}$) fragment being produced, or roughly $10^{4}$ solar-mass fragments, depending on the initial conditions of the flow. The CMB temperature floor is also a key factor in determining the mass of fragments, as it represents a strict temperature lower limit of radiatively cooling gas. Roughly speaking, this scenario occuring at higher redshifts would still demonstrate similar behavior in terms of a critical metallicity, but would produce fragments of slightly higher mass. Overall, this effect is of little importance as even low-redshift star forming molecular clouds do not generally cool below $10\, \text{K}$, wheres as the CMB temperature at $z=10$ is $\sim 30\text\,{K}$. Any observational effects from this temperature floor wold be difficult to extract from our simulations.
\citet{BAI10} studied CMB regulated star formation in some detail (at lower redshifts than we examine) and find that observations currently cannot confirm or refute CMB regulation on star formation.

We did not include dust, even though it can have a major impact in that it allows very efficent $\textrm{H}_2$ formation on grain surfaces, though we do argue its presence will not significantly alter our findings. The presence of dust and its effect on cooling has been discussed as a possible catalyst
for the star formation mode transition. \citet{SCH06} and \citet{SO09} found that
dust induced fragmentation can decrease the Pop II fragmentation mass
scale by about four orders of magnitude over that accessible with metal fine structure lines only. 
\citet{JAP09} suggested that metallicity does not affect the
fragmentation of gas with hot initial conditions, similar to the environment probed in this study. We attribute the difference in our conclusions to our examination of a slightly different physical regime and to our
inclusion of a soft UV background that stifles
$\textrm{H}_2$ cooling \citep[as supported, e.g., by][]{SHA09}. Taking guidance from galaxy formation simulations, we follow isobarically evolving gas in an expanding shocked slab to study its
thermal evolution and fragmentation. Our model is designed to reflect the filament-dominated inflow topology of cold accretion flows, whereas \citet{JAP09} examine the collapse into a spherically-symmetric, centrally concentrated halo potential. In our isobarically evolving case, the tight coupling between density and temperature will increase the importance of the role of the cooling rate in determining the final outcome of fragmentation. This aside,
if we include molecule formation and molecular cooling ($\textrm{H}_2$ and HD) by setting $J_{21} = 10$, our results agree with \citet{Jappsen:09a} in that metal line cooling does not affect the
fragmentation properties of the evolving gas. This emphasizes the
importance of a LW soft UV background in modeling
the formation of Pop~II stars in the first galaxies.

The first galaxies are suspected
to be the first structures where turbulence begins to play a major
role, driven by high Mach-number cold streams penetrating to the
protogalactic core \citep{GB08, WA07}. It is the interplay of
turbulence and self gravity which is suspected to be responsible for
the shape of the present-day IMF, especially the high mass power law extension (Padoan
\& Nordlund 2002; Mac Low \& Klessen 2004; McKee \& Ostriker 2007).
Thus, it is likely that star formation inside the first galaxies will
be characterized by a stellar mass distribution that for the first time
in the history of the universe resembles already the
present-day IMF. Specifically, we can expect turbulence to introduce
density perturbations on the order of ${\cal M}^{2}$, suggesting within our above model that fragments as low as $\sim0.3 \,M_\odot$ might form. Such stars, below the mass limit of $0.8
\,M_\odot$ required to survive for the age of the universe, would still exist today, and could be probed by Stellar Archaeology \citep[][and references therein]{FRE09}. To predict the detailed shape of the Pop. II IMF inside the first galaxies, the source and impact of the turbulence needs to be studied with realistic numerical simulations.

Here, we have constrained some of the key physical processes that govern
star formation in high-redshift protogalaxies within an idealized
analytical framework that allowed us to explore the relevant parameter 
space. This work needs to be followed up by detailed simulations, including
a number of ingredients ignored here, such as
high optical depth effects, H and He ionizing radiation \citep{RIC02}, turbulence, feedback effects (AGN, stellar outflows, etc.), and possibly cosmic rays from the first SNe \citep{JCE07, SB07}. It needs to be emphasized that the estimates presented in this paper are derived with the Bonnor-Ebert mass and a simplified treatment of the post-shock dynamics. To truly probe the fragmentation mass scale in the first galaxies one must use multidimensional numerical simulations. The goal of understanding the complex first stages of galaxy formation seems now within reach, using state-of-the-art three dimensional radiation-hydrodynamical simulations, together with realistic cosmological initial conditions. Such calculations are crucial to make useful predictions for the {\it James Webb Space Telescope (JWST)}.

\acknowledgments

This work was supported in part by NSF grant AST-0708795,  and by NASA ATFP grant
NNX08AL43G (for V.\,B.). The authors thank Farah Prasla and Christopher Lindner for useful conversations and technical help. We thank Sean Couch for adding deuterium chemistry. We also thank Kazu Omukai and Umberto Maio for valuable comments. C.\,S.\,S. thanks L.\,P.\,D for her never-ending patience. This research has made use of NASA's Astrophysics Data System.

\clearpage

\begin{center}
\begin{longtable}{|llll|}
\caption[Reaction Rates for our Chemical Network]{Reaction Rates for our Chemical Network.} \label{tab:chem_reactions} \\

\hline \multicolumn{1}{|c|}{\textbf{Reaction}} & \multicolumn{1}{c|}{\textbf{Rate Coefficient ($\text{cm}^{3}\,\text{s}^{-1}$) }} & \multicolumn{1}{c|}{\textbf{Notes}} & \multicolumn{1}{c|}{\textbf{Reference}} \\ \hline 
\endfirsthead

\multicolumn{4}{c}%
{{\bfseries \tablename\ \thetable{} -- continued from previous page}} \\
\hline \multicolumn{1}{|c|}{\textbf{Reaction}} &
\multicolumn{1}{c|}{\textbf{Rate Coefficient ($\text{cm}^{3}\,\text{s}^{-1}$) }} &
\multicolumn{1}{c|}{\textbf{Notes}} &
\multicolumn{1}{c|}{\textbf{Reference}} \\ \hline 
\endhead

\hline \multicolumn{3}{|r|}{{Continued on next page}} \\ \hline
\endfoot

\hline \hline
\endlastfoot

$\text{H} + \text{e}^{-} \rightarrow\text{H}^{-} + \gamma$   & \begin{tabular}{ l l }\\
\(k_{1}\) & \(=  \text{dex}[-17.845 + 0.762\,\text{log}\,T + 0.1523(\text{log}\,T)^{2}  \) \\
 & \(-\,\, 0.03274(\text{log}\,T)^{3}]  \) \\
 & \(= \text{dex}[-16.420 + 0.1998(\text{log}\,T)^{2} \) \\
 & \(-\,\, 5.447 \times 10^{-3}(\text{log}\,T)^{4} \) \\
 & \(+ \,\,4.0415\times10^{-5}(\text{log}\,T)^{6}] \) \\
 \\
  \end{tabular}
 &\begin{tabular}{ l l }
$T < 6,000\,\text{K}$ \\
 $T > 6,000\, \text{K}$ \\
\\
 \end{tabular}
&  \\

\hline

$\text{H}^{-} + \text{H} \rightarrow\text{H}_{2} + \text{e}^{-}$  &
\begin{tabular}{ l l }\\

\(k_{2}\) & \(=  1.5\times10^{-9} \) \\
 & \(= 4.0\times10^{-9}T^{-0.17} \) \\
 \\
  \end{tabular}
&
\begin{tabular}{ l l }
$T < 300\,\text{K}$ \\
 $T > 300\, \text{K}$ \\
 \end{tabular}
  &  \\

\hline

      $\text{H} + \text{H}^{+} \rightarrow\text{H}_{2}^{+} + \gamma$  &
\begin{tabular}{ l l }\\

\(k_{3}\) & \(=  \text{dex}[-19.38-1.523\,\text{log}\,T \) \\
 & \(+ \,\,1.118(\text{log}\,T)^{2} - 0.1269(\text{log}\,T)^{3}] \) \\
 \\
  \end{tabular}
&
\begin{tabular}{ l l }
 \\
 \\
 \end{tabular}
  &  \\
   \hline

   
   $\text{H} + \text{H}_{2}^{+} \rightarrow\text{H}_{2} + \text{H}^{+}$  &
\begin{tabular}{ l l }\\

\(k_{4}\) & \(=  6.4\times10^{-10} \) \\
\\
  \end{tabular}
&
\begin{tabular}{ l l }
 \\
 \\
 \end{tabular}
  &  \\
   \hline
   

$\text{H}^{-} + \text{H}^{+} \rightarrow\text{H} + \text{H}$  &
\begin{tabular}{ l l }\\

\(k_{5}\) & \(=  5.7\times10^{-6}T^{-0.5} + 6.3\times10^{-8} \) \\
 & \(-\,\,9.2\times10^{-11}T^{0.5} + 4.4\times10^{-13}T \) \\
 \\
  \end{tabular}
&
\begin{tabular}{ l l }
 \\
 \\
 \end{tabular}
  &  \\
   \hline
   

   $\text{H}^{+}_{2} + \text{e}^{-} \rightarrow\text{H} + \text{H}$  &
\begin{tabular}{ l l }\\

\(k_{6}\) & \(= 1.0\times10^{-8} \) \\
 & \(=1.32\times10^{-6}T^{-0.76} \) \\
 \\
  \end{tabular}
&
\begin{tabular}{ l l }
$T < 617\,\text{K}$ \\
 $T > 617\, \text{K}$ \\
 \end{tabular}
  &  \\
   \hline


   $\text{H}_{2} + \text{H}^{+} \rightarrow\text{H}_{2}^{+} + \text{H}$  &
\begin{tabular}{ l l }\\
\(k_{7}\) & \(= [-3.3232183\times10^{-7} \) \\
 & \(\,\,\,\,\,\,\,\,\,+\,3.3735382\times10^{-7}\text{ln}\,T \) \\
  & \(\,\,\,\,\,\,\,\,\,-\,1.4491368\times10^{-7}(\text{ln}\,T)^{2} \) \\
   & \(\,\,\,\,\,\,\,\,\,+\,3.4172805\times10^{-8}(\text{ln}\,T)^{3} \) \\
    & \(\,\,\,\,\,\,\,\,\,-\,4.7813720\times10^{-9}(\text{ln}\,T)^{4} \) \\
     & \(\,\,\,\,\,\,\,\,\,+\,3.9731542\times10^{-10}(\text{ln}\,T)^{5} \) \\
      & \(\,\,\,\,\,\,\,\,\,-\,1.8171411\times10^{-11}(\text{ln}\,T)^{6} \) \\
        & \(\,\,\,\,\,\,\,\,\,+\,3.5311932\times10^{-13}(\text{ln}\,T)^{7}] \) \\
      & \(\,\,\,\,\,\,\,\,\,\times\,\text{exp}(\frac{-21237.15}{T}) \) \\
      \\
  \end{tabular}
&
\begin{tabular}{ l l }
 \\
  \\
 \end{tabular}
  &  \\
   \hline


   $\text{H}_{2} + \text{e}^{-} \rightarrow\text{H} + \text{H} + \text{e}^{-}$  &
\begin{tabular}{ l l }\\
\(k_{8}\) & \(= 3.73\times10^{-9}T^{0.1121}\text{exp}(\frac{-99430}{T}) \) \\
 \\
  \end{tabular}
&
\begin{tabular}{ l l }
 \\
  \\
 \end{tabular}
  &  \\
   \hline


   $\text{H}_{2} + \text{H} \rightarrow\text{H} + \text{H} + \text{H}$  &
\begin{tabular}{ l l }\\

\(k_{9, \text{l}}\) & \(= 6.67\times10^{-12}T^{0.5}\text{exp}[-(1.0\, + \frac{63590}{T})] \) \\
\\
\(k_{9, \text{h}}\) & \(= 3.52\times10^{-9}\text{exp}(-\frac{43900}{T})\) \\
\\
\(n_{\text{cr}, \text{H}}\) & \(= \text{dex}\left[4.0\, - \,0.416\,\text{log}\,(\frac{T}{10000}) \,-\, 0.327\,\text{log}^{2}(\frac{T}{10000})\right] \) \\
 \\
  \end{tabular}
&
\begin{tabular}{ l l }
 \\
  \\
 \end{tabular}
  &  \\
   \hline


   $\text{H}_{2} + \text{H}_{2} \rightarrow\text{H}_{2} + \text{H} + \text{H}$  &
\begin{tabular}{ l l }\\

\(k_{10, \text{l}}\) & \(= \frac{5.996\times10^{-30}T^{4.1881}}{(1.0+6.761\times10^{-6}T)^{5.6881}}\text{exp}\,(-\frac{54657.4}{T})\) \\
\\
\(k_{10, \text{h}}\) & \(= 1.3\times10^{-9}\text{exp}\,(-\frac{53300}{T}) \) \\
\\
\(n_{\text{cr}, \text{H}_{2}}\) & \(= \text{dex}\left[4.845 - 1.3\,\text{log}\,(\frac{T}{10000}) + 1.62\,\text{log}^{2}(\frac{T}{10000})\right] \) \\
 \\
  \end{tabular}
&
\begin{tabular}{ l l }
 \\
  \\
 \end{tabular}
  &  \\
   \hline


   $\text{H} + \text{e}^{-} \rightarrow\text{H}^{+} + \text{e}^{-}+ \text{e}^{-}$  &
\begin{tabular}{ l l }\\
\(k_{11}\) & \(= \text{exp}[-3.271396786\times10 \) \\
 & \(\,\,\,\,\,\,\,\,\,+\,1.35365560\times10\,\text{ln}\,T_{\text{e}} \) \\
  & \(\,\,\,\,\,\,\,\,\,-\,5.73932875\times10^{0}\,(\text{ln}\,T_{\text{e}})^{2} \) \\
   & \(\,\,\,\,\,\,\,\,\,+\,1.56315498\times10^{0}\,(\text{ln}\,T_{\text{e}})^{3} \) \\
    & \(\,\,\,\,\,\,\,\,\,-\,2.87705600\times10^{-1}\,(\text{ln}\,T_{\text{e}})^{4} \) \\
     & \(\,\,\,\,\,\,\,\,\,+\,3.48255977\times10^{-2}\,(\text{ln}\,T_{\text{e}})^{5} \) \\
      & \(\,\,\,\,\,\,\,\,\,-\,2.63197617\times10^{-3}(\text{ln}\,T_{\text{e}})^{6} \) \\
        & \(\,\,\,\,\,\,\,\,\,+\,1.11954395\times10^{-4}(\text{ln}\,T_{\text{e}})^{7} \) \\
        & \(\,\,\,\,\,\,\,\,\,-\,2.03914985\times10^{-6}(\text{ln}\,T_{\text{e}})^{8}] \) \\
      \\
  \end{tabular}
&
\begin{tabular}{ l l }
 \\
  \\
 \end{tabular}
  &  \\
   \hline


   $\text{D} + \text{e}^{-} \rightarrow\text{D}^{+} + \text{e}^{-} + \text{e}^{-}$  &
\begin{tabular}{ l l }\\
\(k_{12}\) & \(= k_{11} \) \\
 \\
  \end{tabular}
&
\begin{tabular}{ l l }
 \\
  \\
 \end{tabular}
  &  \\
   \hline


   $\text{H}^{+} + \text{e}^{-} \rightarrow\text{H} + \gamma$  &
\begin{tabular}{ l l }\\
\(k_{13}\) & \(= 2.753 \times10^{-14}(\frac{315614}{T})^{1.503} \) \\
 & \(\times[1.0 + (\frac{604625}{T})^{0.470}]^{-1.923} \) \\
 \\
  \end{tabular}
&
\begin{tabular}{ l l }
 \\
  \\
 \end{tabular}
  &  \\
   \hline


      $\text{D}^{+} + \text{e}^{-} \rightarrow\text{D} + \gamma$  &
\begin{tabular}{ l l }\\
\(k_{14}\) & \(= k_{13} \) \\
 \\
  \end{tabular}
&
\begin{tabular}{ l l }
 \\
  \\
 \end{tabular}
  &  \\
   \hline


      $\text{H}^{-} + \text{e}^{-} \rightarrow\text{H} + \text{e}^{-}+ \text{e}^{-}$  &
\begin{tabular}{ l l }\\
\(k_{15}\) & \(= \text{exp}[-1.801849334\times10 \) \\
 & \(\,\,\,\,\,\,\,\,\,+\,2.36085220\times10^{0}\,\text{ln}\,T_{\text{e}} \) \\
  & \(\,\,\,\,\,\,\,\,\,-\,2.82744300\times10^{-1}\,(\text{ln}\,T_{\text{e}})^{2} \) \\
   & \(\,\,\,\,\,\,\,\,\,+\,1.62331664\times10^{-2}\,(\text{ln}\,T_{\text{e}})^{3} \) \\
    & \(\,\,\,\,\,\,\,\,\,-\,3.36501203\times10^{-2}\,(\text{ln}\,T_{\text{e}})^{4} \) \\
     & \(\,\,\,\,\,\,\,\,\,+\,1.17832978\times10^{-2}\,(\text{ln}\,T_{\text{e}})^{5} \) \\
      & \(\,\,\,\,\,\,\,\,\,-\,1.65619470\times10^{-3}(\text{ln}\,T_{\text{e}})^{6} \) \\
        & \(\,\,\,\,\,\,\,\,\,+\,1.06827520\times10^{-4}(\text{ln}\,T_{\text{e}})^{7} \) \\
        & \(\,\,\,\,\,\,\,\,\,-\,2.63128581\times10^{-6}(\text{ln}\,T_{\text{e}})^{8}] \) \\
      \\
  \end{tabular}
&
\begin{tabular}{ l l }
 \\
  \\
 \end{tabular}
  &  \\
   \hline


      $\text{H}^{-} + \text{H} \rightarrow\text{H} + \text{H}+ \text{e}^{-}$  &
\begin{tabular}{ l l }\\
\(k_{16}\) & \(= 2.5634\times10^{-9}T_{\text{e}}^{1.78186} \) \\
 & \( = \text{exp}[-2.0372609\times10 \) \\
  & \(\,\,\,\,\,\,\,\,\,+\,1.13944933\times10^{0}\,(\text{ln}\,T_{\text{e}}) \) \\
   & \(\,\,\,\,\,\,\,\,\,-\,1.4210135\times10^{-1}\,(\text{ln}\,T_{\text{e}})^{2} \) \\
    & \(\,\,\,\,\,\,\,\,\,+\,8.4644554\times10^{-3}\,(\text{ln}\,T_{\text{e}})^{3} \) \\
     & \(\,\,\,\,\,\,\,\,\,-\,1.4327641\times10^{-3}\,(\text{ln}\,T_{\text{e}})^{4} \) \\
      & \(\,\,\,\,\,\,\,\,\,+\,2.0122503\times10^{-4}(\text{ln}\,T_{\text{e}})^{5} \) \\
        & \(\,\,\,\,\,\,\,\,\,+\,8.6639632\times10^{-5}(\text{ln}\,T_{\text{e}})^{6} \) \\
        & \(\,\,\,\,\,\,\,\,\,-\,2.5850097\times10^{-5}(\text{ln}\,T_{\text{e}})^{7} \) \\
        & \(\,\,\,\,\,\,\,\,\,+\,2.4555012\times10^{-6}(\text{ln}\,T_{\text{e}})^{8} \) \\
        & \(\,\,\,\,\,\,\,\,\,-\,8.0683825\times10^{-8}(\text{ln}\,T_{\text{e}})^{9}] \) \\
      \\
  \end{tabular}
&
\begin{tabular}{ l l }
$T_{\text{e}} < 0.1\,\text{eV}$ \\
\\
\\
\\
\\
\\
\\
\\
\\
\\
\\
 $T_{\text{e}} > 0.1\, \text{eV}$ \\
 \end{tabular}
  &  \\
   \hline


      $\text{H}^{-} + \text{H}^{+} \rightarrow\text{H}_{2}^{+} +  \text{e}^{-}$  &
\begin{tabular}{ l l }\\
\(k_{17}\) & \(= 6.9\times10^{-9}\,T^{-0.35} \) \\
  & \( = 9.6\times10^{-7}T^{-0.90} \) \\
      \\
  \end{tabular}
&
\begin{tabular}{ l l }
$T< 8000\,\text{K}$ \\

 $T > 8000\, \text{K}$ \\
 \end{tabular}
  &  \\
   \hline


      $\text{H} + \text{D}^{+} \rightarrow\text{D} +  \text{H}^{+}$  &
\begin{tabular}{ l l }\\
\(k_{18}\) & \(= 2.06\times10^{-10}T^{0.396}\text{exp}(-\frac{33}{T}) \) \\
  & \( + \,\,2.03\times10^{-9}T^{-0.332} \) \\
      \\
  \end{tabular}
&
\begin{tabular}{ l l }

 \end{tabular}
  &  \\
   \hline


         $\text{H}^{+} + \text{D} \rightarrow\text{H} +  \text{D}^{+}$  &
\begin{tabular}{ l l }\\
\(k_{19}\) & \(= 2.0\times10^{-10}T^{0.402}\text{exp}(-\frac{37.1}{T}) \) \\
  & \( -\,\,3.31\times10^{-17}T^{1.48} \) \\
    & \( = 3.44\times10^{-10}T^{0.35} \) \\
      \\
  \end{tabular}
&
\begin{tabular}{ l l }
$T< 2\times10^{5}\,\text{K}$ \\
\\

 $T > 2\times10^{5}\, \text{K}$ \\
 \end{tabular}
  &  \\
   \hline


            $\text{H}_{2} + \text{D}^{+} \rightarrow\text{HD} +  \text{H}^{+}$  &
\begin{tabular}{ l l }\\
\(k_{20}\) & \(= [0.417 \,+ \,0.846\,\text{log}\,T\,-\,0.137\,(\text{log}\,T)^{2}]\times10^{-9} \) \\

      \\
  \end{tabular}
&
\begin{tabular}{ l l }

 \end{tabular}
  &  \\
   \hline


  $\text{HD} + \text{H}^{+} \rightarrow\text{H}_{2} +  \text{D}^{+}$  &
\begin{tabular}{ l l }\\
\(k_{21}\) & \(= 1.1\times10^{-9}\,\text{exp}\,(-\frac{488}{T}) \) \\
\\
  \end{tabular}
&
\begin{tabular}{ l l }

 \end{tabular}
  &  \\
   \hline


  $\text{D} + \text{H}_{2} \rightarrow\text{HD} +  \text{H}$  &
\begin{tabular}{ l l }\\
\(k_{22}\) & \(= 1.69\times10^{-10}\,\text{exp}\,(-\frac{4680}{T}) \) \\
  & \( = 1.69\times10^{-10}\,\text{exp}\,(-\frac{4680}{T} + \frac{198800}{T^{2}}) \) \\
\\
  \end{tabular}
&
\begin{tabular}{ l l }
$T< 200\,\text{K}$ \\

 $T > 200\, \text{K}$ \\
 \end{tabular}
  &  \\
   \hline

   
     $\text{HD} + \text{H} \rightarrow\text{D} +  \text{H}_{2}$  &
\begin{tabular}{ l l }\\
\(k_{23}\) & \(= 5.25\times10^{-11}\,\text{exp}\,(-\frac{4430}{T}) \) \\
  & \( = 5.25\times10^{-11}\,\text{exp}\,(-\frac{4430}{T} + \frac{173900}{T^{2}}) \) \\
\\
  \end{tabular}
&
\begin{tabular}{ l l }
$T< 200\,\text{K}$ \\

 $T > 200\, \text{K}$ \\
 \end{tabular}
  &  \\
   \hline


   $\text{He} + \text{e}^{-} \rightarrow\text{He}^{+} +  \text{e}^{-} + \text{e}^{-}$  &
\begin{tabular}{ l l }\\
\(k_{24}\) & \(= \text{exp}\,[-4.409864886 \times 10 \) \\
  & \( \,\,\,\,\,\,\,\,\,+\,2.391596563 \times10\,\text{ln}\,T_{\text{e}} \) \\
    & \( \,\,\,\,\,\,\,\,\,-\,1.07532302\times10\,(\text{ln}\,T_{\text{e}})^{2} \) \\
    & \( \,\,\,\,\,\,\,\,\,+\,3.05803875\times10^{0}\,(\text{ln}\,T_{\text{e}})^{3} \) \\
    & \( \,\,\,\,\,\,\,\,\,-\,5.6851189\times10^{-1}\,(\text{ln}\,T_{\text{e}})^{4} \) \\
    & \( \,\,\,\,\,\,\,\,\,+\,6.79539123\times10^{-2}\,(\text{ln}\,T_{\text{e}})^{5} \) \\
    & \( \,\,\,\,\,\,\,\,\,-\,5.0090561\times10^{-3}\,(\text{ln}\,T_{\text{e}})^{6} \) \\
    & \( \,\,\,\,\,\,\,\,\,+\,2.06723616\times10^{-4}\,(\text{ln}\,T_{\text{e}})^{7} \) \\
    & \( \,\,\,\,\,\,\,\,\,-\,3.64916141\times10^{-6}\,(\text{ln}\,T_{\text{e}})^{8}] \) \\
\\
  \end{tabular}
&
\begin{tabular}{ l l }

 \end{tabular}
  &  \\
   \hline


     $\text{He}^{+} + \text{e}^{-} \rightarrow\text{He} +  \gamma$  &
\begin{tabular}{ l l }\\
\(k_{25, \text{rr}}\) & \(= 10^{-11}\,T^{-0.5}\,[11.19\,-\,1.676\,\text{log}\,T \) \\
  & \(\,\,\,\,\,\,\, -\,0.2852\,(\text{log}\,T)^{2}\,+\,0.04433\,(\text{log}\,T)^{3}] \) \\
\\
\(k_{25, \text{di}}\) & \(= 1.9\times10^{-3}\,T^{-1.5}\,\text{exp}\,(-\frac{473421}{T}) \) \\
  & \(\,\,\,\,\,\,\, \times\,[1.0\, + \,0.3\,\,\text{exp}\,(-\frac{94684}{T})] \) \\
  \\
  \end{tabular}
&
\begin{tabular}{ l l }

 \end{tabular}
  &  \\
   \hline


        $\text{He}^{+} + \text{H} \rightarrow\text{He} + \text{H}^{+}$  &
\begin{tabular}{ l l }\\
\(k_{26}\) & \(= 1.25\times10^{-15}(\frac{T}{300})^{0.25} \) \\
\\
  \end{tabular}
&
\begin{tabular}{ l l }

 \end{tabular}
  &  \\
   \hline


     $\text{He} + \text{H}^{+} \rightarrow\text{He}^{+} +  \text{H}$  &
\begin{tabular}{ l l }\\
\(k_{27}\) & \(= 1.26\times10^{-9}\,T^{-0.75}\text{exp}\,(-\frac{127500}{T}) \) \\
  & \( = 4.0\times10^{-37}\,T^{4.74}\) \\
\\
  \end{tabular}
&
\begin{tabular}{ l l }
$T< 10^{4}\,\text{K}$ \\

 $T > 10^{4}\, \text{K}$ \\
 \end{tabular}
  &  \\
   \hline


  $\text{He}^{+} + \text{D} \rightarrow\text{He} +  \text{D}^{+}$  &
\begin{tabular}{ l l }\\
\(k_{28}\) & \(= k_{26} \) \\

\\
  \end{tabular}
&
\begin{tabular}{ l l }

 \end{tabular}
  &  \\
   \hline
   

     $\text{He} + \text{D}^{+} \rightarrow\text{He}^{+} +  \text{D}$  &
\begin{tabular}{ l l }\\
\(k_{29}\) & \(= k_{27} \) \\

\\
  \end{tabular}
&
\begin{tabular}{ l l }

 \end{tabular}
  &  \\
   \hline


$\text{C}^{+} + \text{e}^{-} \rightarrow\text{C} + \gamma$  &
\begin{tabular}{ l l }\\

\(k_{30}\) & \(=  4.67 \times 10^{-12} \left(\frac{T}{300}\right)^{-0.6} \) \\
 & \(= 1.23 \times 10 ^{-17} \left(\frac{T}{300}\right)^{2.49}\text{exp}\left(\frac{21845.6}{T}\right)  \) \\
 & \(= 9.62 \times 10^{-8} \left(\frac{T}{300}\right)^{-1.37}\text{exp}\left(\frac{-115786.2}{T}\right) \) \\
 \\
  \end{tabular}
&
\begin{tabular}{ l l }
T $<$ 7950\,\text{K}  \\
7950 K $<$ T $<$ 21140 \,\text{K}   \\
T $>$ 21140\, \text{K} \\
 \end{tabular}
  &  \\
   \hline

 
   $\text{Si}^{+} + \text{e}^{-} \rightarrow \text{Si} + \gamma$  &
\begin{tabular}{ l l }\\
\(k_{31}\) & \(=  7.5 \times 10^{-12} \left(\frac{T}{300}\right)^{-0.55} \) \\
 & \(= 4.86 \times 10 ^{-12} \left(\frac{T}{300}\right)^{-0.32}  \) \\
 & \(= 9.08 \times 10^{-14} \left(\frac{T}{300}\right)^{0.818} \) \\
 \\
  \end{tabular}
&
\begin{tabular}{ l l }
$T < 2000$\,\text{K}  \\
$2,000 K < T < 10,000$ \,\text{K}   \\
$T > 10,000$\, \text{K} \\
 \end{tabular}
  &  \\
   \hline
 

   $\text{O}^{+} + \text{e}^{-} \rightarrow\text{O} + \gamma$  &
\begin{tabular}{ l l }\\
\(k_{32}\) & \(=  1.30 \times 10^{-10}T^{-0.64} \) \\
 & \(= 1.41 \times 10 ^{-10}T^{-0.66} + 7.4 \times 10^{-4}T^{-1.5} \) \\
 & \(\times \, \, \text{exp}\left(-\frac{175000}{T}\right)[1.0 \,+\, 0.062 \,\text{exp}\left(-\frac{145000}{T}\right)] \) \\
 \\
  \end{tabular}
&
\begin{tabular}{ l l }
$T < 400$\,\text{K}  \\
  \\
$T > 400$\, \text{K} \\
 \end{tabular}
  &  \\
   \hline
 
 
 $\text{C} + \text{e}^{-} \rightarrow\text{C}^{+} + \text{e}^{-} + \text{e}^{-} $  &
\begin{tabular}{ r l }\\
\(k_{33}\) & \(=  6.85 \times 10^{-8} (0.193 + u)^{-1}u^{0.25}e^{-u} \) \\
\\
  \end{tabular}
&
\begin{tabular}{ l l }
$u = 11.26 / T_{\text{e}}$
 \end{tabular}
  &  \\
   \hline
 

    $\text{Si} + \text{e}^{-} \rightarrow\text{Si}^{+} + \text{e}^{-} + \text{e}^{-} $  &
\begin{tabular}{ r l }\\
\(k_{34}\) & \(=  1.88 \times 10^{-7} (1.0 + u^{0.5})(0.376 + u)^{-1}u^{0.25}e^{-u} \) \\
\\
  \end{tabular}
&
\begin{tabular}{ l l }
$u = 8.2 / T_{\text{e}}$
 \end{tabular}
  &  \\
   \hline
 
 
       $\text{O} + \text{e}^{-} \rightarrow\text{O}^{+} + \text{e}^{-} + \text{e}^{-} $  &
\begin{tabular}{ r l }\\
\(k_{35}\) & \(= 3.59 \times 10^{-8} (0.073 + u)^{-1}u^{0.34}e^{-u} \) \\
\\
  \end{tabular}
&
\begin{tabular}{ l l }
$u = 13.6 / T_{\text{e}}$
 \end{tabular}
  &  \\
   \hline
 
 
 $\text{O}^{+} + \text{H} \rightarrow\text{O} + \text{H}^{+} $  &
\begin{tabular}{ r l }\\
\(k_{36}\) & \(= 4.99 \times 10^{-11} T^{0.405} + 7.54 \times 10^{-10}T^{-0.458} \) \\
\\
  \end{tabular}
&
&  \\
 \hline
 

  $\text{O} + \text{H}^{+} \rightarrow\text{O}^{+} + \text{H} $  &
\begin{tabular}{ r l }\\
\(k_{37}\) & \(= [1.08 \times 10^{-11}T^{0.517}  \) \\
& \( \,\,\,\,\,\,+ 4.00 \times 10^{-10}T^{0.00669}]\,\text{exp}(-\frac{227}{T})   \) \\
\\
  \end{tabular}
&
&  \\
 \hline

 
  $\text{O} + \text{He}^{+} \rightarrow\text{O}^{+} + \text{He} $  &
\begin{tabular}{ r l }\\
\(k_{38}\) & \(= 4.991 \times 10^{-15}(\frac{T}{10000})^{0.3794}\text{exp}(-\frac{T}{1121000})  \) \\
& \( \,\,\,\,\,\,+ 2.780 \times 10^{-15}(\frac{T}{10000})^{-0.2163}\text{exp}(\frac{T}{815800})   \) \\
\\
  \end{tabular}
&
&  \\
 \hline
 
 
    $\text{C} + \text{H}^{+} \rightarrow\text{C}^{+} + \text{H} $  &
\begin{tabular}{ r l }\\
\(k_{39}\) & \(= 3.9\times10^{-16}T^{0.213} \) \\
\\
  \end{tabular}
&
&  \\
 \hline
 
 
   $\text{C} + \text{H}^{+} \rightarrow\text{C}^{+} + \text{H} $  &
\begin{tabular}{ r l }\\
\(k_{40}\) & \(= 6.08\times10^{-14}(\frac{T}{10000})^{1.96}\text{exp}(-\frac{170000}{T})\) \\
\\
  \end{tabular}
&
&  \\
 \hline


   $\text{C} + \text{He}^{+} \rightarrow\text{C}^{+} + \text{He}$  &
\begin{tabular}{ r l }\\
\(k_{41}\) & \(=  8.58\times10^{-17}T^{0.757} \) \\
 & \(= 3.25\times10^{-17}T^{0.968} \) \\
 & \(= 2.77\times10^{-19}T^{1.597} \) \\
 \\
  \end{tabular}
&
\begin{tabular}{ l l }
$T \leq 200\,\text{K} $ \\
$200 K < T \leq 2000 \,\text{K} $  \\
$T > 2000\, \text{K}$ \\
 \end{tabular}
  &  \\
   \hline
 
 
      $\text{Si} + \text{H}^{+} \rightarrow\text{Si}^{+} + \text{H}$  &
\begin{tabular}{ r l }\\
\(k_{42}\) & \(=  5.88\times10^{-13}T^{0.848} \) \\
 & \(= 1.45\times10^{-13}T \) \\
 \\

  \end{tabular}
&
\begin{tabular}{ l l }
$T \leq 10^{4}\,\text{K}$  \\
$T > 10^{4}\, \text{K}$ \\
 \end{tabular}
  &  \\
   \hline 
 
 
   $\text{Si} + \text{He}^{+} \rightarrow\text{Si}^{+} + \text{He} $  &
\begin{tabular}{ r l }\\
\(k_{43}\) & \(= 3.3\times10^{-9} \) \\
\\
  \end{tabular}
&
&  \\
 \hline
 

   $\text{C}^{+} + \text{Si} \rightarrow\text{C} + \text{Si}^{+} $  &
\begin{tabular}{ r l }\\
\(k_{44}\) & \(= 2.1\times10^{-9} \) \\
\\
  \end{tabular}
&
&  \\

\hline


   $\text{H}_{2} + \text{He} \rightarrow\text{H} + \text{H} + \text{He} $  &
\begin{tabular}{ l l }\\
\(k_{75, \text{l}}\) & \(= \text{dex}\,[-27.029 \,+\, 3.801\,\text{log}\,T\,-\,(\frac{29487}{T})] \) \\
\\
\(k_{75, \text{h}}\) & \(= \text{dex}\,[-2.729 \,-\, 1.75\,\text{log}\,T\,-\,(\frac{23474}{T})] \) \\
\\
\(n_{\text{cr}, \text{He}}\) & \(= \text{dex}\,[5.0792 - 6.247\times10^{-5}\,(T \,-\, 2000)] \) \\
\\
  \end{tabular}
&
&  \\
 \hline


    $\text{H}_{2} + \text{He}^{+} \rightarrow\text{He} + \text{H}_{2}^{+} $  &
\begin{tabular}{ r l }\\
\(k_{138}\) & \(= 7.2\times10^{-15} \) \\
\\
  \end{tabular}
&
&  \\
 \hline
 
 
   $\text{H}_{2} + \text{He}^{+} \rightarrow\text{He} + \text{H} + \text{H}^{+} $  &
\begin{tabular}{ r l }\\
\(k_{139}\) & \(= 3.7\times10^{-14}\,\text{exp}\,(-\frac{35}{T}) \) \\
\\
  \end{tabular}
&
&  \\
 \hline


   $\text{H} + \text{H} + \text{H} \rightarrow\text{H}_{2} + \text{H} $  &
\begin{tabular}{ r l }\\
\(k_{206}\) & \(= 1.32\times10^{-32}\,(\frac{T}{300})^{-0.38} \) \\
 & \(= 1.32\times10^{-32}\,(\frac{T}{300})^{-1.0} \) \\
\\
  \end{tabular}
&
\begin{tabular}{ l l }
$T \leq 300\,\text{K}$  \\
$T > 300\, \text{K}$ \\
 \end{tabular}
&  \\
 \hline

 
  $\text{H} + \text{H} + \text{H}_{2} \rightarrow\text{H}_{2} + \text{H}_{2} $  &
\begin{tabular}{ r l }\\
\(k_{207}\) & \(= 2.8\times10^{-31}\,T^{-0.6} \) \\

\\
  \end{tabular}
&
\begin{tabular}{ l l }

 \end{tabular}
&  \\
 \hline

 
  $\text{H} + \text{H} + \text{He} \rightarrow\text{H}_{2} + \text{He} $  &
\begin{tabular}{ r l }\\
\(k_{208}\) & \(= 6.9\times10^{-32}\,T^{-0.4} \) \\

\\
  \end{tabular}
&
\begin{tabular}{ l l }

 \end{tabular}
&  \\
 \hline
 
 
   $\text{Fe} + \text{e}^{-}  \rightarrow\text{Fe}_{+} + \text{e}^{-} + \text{e}^{-} $  &
\begin{tabular}{ r l }\\
\(k_{222}\) & \(= 2.52\times10^{-7}\,(0.701\,+\,u)^{-1.0}\,(u)^{0.25}\text{exp}(-u) \) \\

\\
  \end{tabular}
&
\begin{tabular}{ l l }
$u = 7.9 / T_{\text{e}}$
 \end{tabular}
& 1 \\
 \hline


    $\text{Fe}^{+} + \text{e}^{-}  \rightarrow\text{Fe} + \gamma$  &
\begin{tabular}{ r l }\\
\(k_{223}\) & \(= 1.42\times10^{-13}\,(\frac{T}{10^{4}})^{-0.891} \) \\
& \( \,\,\,\,\,\,+\, 1.6\times10^{-3}\,T^{-1.5}\,\text{exp}\,(-\frac{6.0\times10^{4}}{T}) \) \\
 &    \( \,\left[1.0 \,+\,0.46\,\text{exp}\, (-\frac{8.97\times10^{4}}{T})\right] \) \\
\\
  \end{tabular}
&
\begin{tabular}{ l l }

 \end{tabular}
& 2 \\
 \hline


   $\text{He}^{+} + \text{e}^{-} \rightarrow\text{He}^{++} + \text{e}^{-} +  \text{e}^{-} $  &
\begin{tabular}{ r l }\\
\(k_{300}\) & \(= 5.68\times10^{-12}\,\text{exp}\,(-\frac{631515}{T})\, \) \\
& \( \,\,\,\,\,\,\times \,(T^{-0.5}\, + \,3.16228\times10^{-3})^{-1}  \) \\

\\
  \end{tabular}
&
\begin{tabular}{ l l }

 \end{tabular}
& 3 \\
 \hline
 
 
  $\text{He}^{++} + \text{e}^{-} \rightarrow\text{He}^{+} +\gamma $  &
\begin{tabular}{ r l }\\
\(k_{301}\) & \(= 1.34\times10^{-9}\,T^{-0.7}\,[1.0 \,+\, (\frac{T}{10^{6}})^{0.7}]^{-1.0}\) \\

\\
  \end{tabular}
&
\begin{tabular}{ l l }

 \end{tabular}
& 3 \\

\end{longtable}
\end{center}


\end{document}